\documentclass[aip,twocolumn]{revtex4}

\usepackage{amsmath}
\usepackage{amssymb}
\usepackage[colorlinks]{hyperref}
\usepackage{color}

\def\ov#1{\overline{#1}}

\def\wt#1{\widetilde{#1}}
\def\vb#1{\mbox{\boldmath$#1$}}
\def\pd#1#2{\frac{\partial #1}{\partial #2}}
\def\fd#1#2{\frac{\delta #1}{\delta #2}}
\def\wh#1{\widehat{#1}}
\def\bdot{\,\vb{\cdot}\,}
\def\btimes{\,\vb{\times}\,}

\def\bhat{\wh{{\sf b}}}
\def\cal#1{\mathcal{#1}}

\def\bhat{\wh{{\sf b}}}
\def\exd{{\sf d}}

\newcommand{\rem}[1]{}

\newcommand{\bc}{\begin{center}}
\newcommand{\ec}{\end{center}}
\newcommand{\bt}{\begin{tabbing}}
\newcommand{\et}{\end{tabbing}}
\newcommand{\be}{\begin{eqnarray*}}
\newcommand{\ee}{\end{eqnarray*}}
\newcommand{\bs}{\begin{slide}}
\newcommand{\es}{\end{slide}}

\begin{document}

\title{Variational principles for the guiding-center Vlasov-Maxwell equations}

\author{Alain J.~Brizard$^{1}$ and Cesare Tronci$^{2}$}
\affiliation{$^{1}$Department of Physics, Saint Michael's College, Colchester, VT 05439, USA \\ $^{2}$Department of Mathematics, University of Surrey, Guildford, GU2 7XH, United Kingdom}

\date{February 15, 2016}

\begin{abstract}
The Lagrange, Euler, and Euler-Poincar\'{e} variational principles for the guiding-center Vlasov-Maxwell equations are presented. Each variational principle presents a different approach to deriving guiding-center polarization and magnetization effects into the guiding-center Maxwell equations. The conservation laws of energy, momentum, and angular momentum are also derived by Noether method, where the guiding-center stress tensor is now shown to be explicitly symmetric.
\end{abstract}


\maketitle

\section{Guiding-center Vlasov-Maxwell Equations}

The guiding-center formulation of charged-particle dynamics in nonuniform magnetized plasmas represents one of the most important paradigms in plasma physics. In particular, the Hamiltonian structure of single-particle guiding-center dynamics has proved tremendously useful \cite{Cary_Brizard_2009} in the theoretical and numerical analysis of magnetically-confined plasmas. Because of the recent development of variational integration numerical techniques
\cite{Squire_Qin_Tang_2012,Krauss_2013,Evstatiev_Shadwick_2013,Evstatiev_2014,Ellison_2015}, it is the primary purpose of the present paper to investigate the variational structures of the guiding-center Vlasov-Maxwell equations.

The guiding-center Vlasov-Maxwell equations describe the coupled time evolution of the guiding-center Vlasov distribution function $f_{\mu}({\bf X},p_{\|},t)$, and the electromagnetic fields ${\bf E}({\bf x},t)$ and ${\bf B}({\bf x},t)$. Here, ${\bf X}$ denotes the guiding-center position, while ${\bf x}$ denotes the field position, $p_{\|}$ denotes the parallel guiding-center (kinetic) momentum, $\mu$ denotes the guiding-center magnetic moment (which is a guiding-center invariant), and the guiding-center gyroangle $\theta$ (which is canonically conjugate to the guiding-center gyroaction $\mu\,B/\Omega$) is an ignorable coordinate  \cite{Cary_Brizard_2009} (i.e., $\partial f_{\mu}/\partial\theta \equiv 0$). The reduced guiding-center phase space is, therefore, four dimensional with coordinates $z^{a} \equiv ({\bf X},p_{\|})$, while $\mu$ appears as a label on the guiding-center Vlasov distribution function $f_{\mu}({\bf z},t)$.

In contrast to the drift-kinetic \cite{Hazeltine_1973} and gyrokinetic \cite{Brizard-Hahm_2007}  Vlasov-Maxwell equations, the electromagnetic fields ${\bf E}({\bf x},t)$ and 
${\bf B}({\bf x},t)$ appearing in the present work are not separated into a time-independent background magnetic field ${\bf B}_{0}({\bf x})$ and time-dependent electromagnetic field perturbations ${\bf E}_{1}({\bf x},t)$ and ${\bf B}_{1}({\bf x},t)$ that satisfy separate space-time orderings from the background magnetic field ${\bf B}_{0}({\bf x})$. Instead, all 
Vlasov-Maxwell fields $(f_{\mu},{\bf E},{\bf B})$ obey the same space-time orderings in which their time dependence is slow compared to the fast gyrofrequency $\Omega = eB/mc$ of each particle species (with mass $m$ and charge $e$) while their weak spatial dependence is used to guarantee the validity of a guiding-center Hamiltonian representation of charged-particle dynamics \cite{Cary_Brizard_2009}.

Moreover, in contrast to previous guiding-center Vlasov-Maxwell theories \cite{Pfirsch_1984,Pfirsch-Morrison_1985,CRPW_1986}, our work does not introduce a transformation to a frame of reference drifting with the $E\times B$ velocity. Hence, the guiding-center symplectic structure and the guiding-center Hamiltonian do not acquire an explicit dependence on the electric field ${\bf E}$ and, thus, the standard polarization does not appear explicitly in our work. As we shall see, however, the guiding-center polarization does appear as the standard moving electric-dipole contribution to the guiding-center magnetization. It is one of the purposes of this paper to show how this moving-dipole contribution emerges naturally from the variational structure of the guiding-center Maxwell-Vlasov system.

\subsection{Guiding-center Vlasov equation}

The guiding-center Vlasov equation for the Vlasov distribution function $f_{\mu}({\bf z},t)$ is expressed as
\begin{equation}
\pd{f_{\mu}}{t} = -\frac{d_{\rm gc}{\bf X}}{dt}\bdot\nabla f_{\mu} - \frac{d_{\rm gc}p_{\|}}{dt}\;\pd{f_{\mu}}{p_{\|}} \equiv -\;\frac{d_{\rm gc}z^{a}}{dt}\;
\pd{f_{\mu}}{z^{a}},
\label{eq:gcV}
\end{equation}
where summation over repeated indices is implied and the guiding-center equations of motion \cite{Littlejohn}:
\begin{eqnarray}
\frac{d_{\rm gc}{\bf X}}{dt} & = & \frac{p_{\|}}{m}\,\frac{{\bf B}^{*}}{B_{\|}^{*}} \;+\; {\bf E}^{*}\btimes\frac{c\,\bhat}{B_{\|}^{*}}, \label{eq:Xgc_dot} \\
\frac{d_{\rm gc}p_{\|}}{dt} & = & e\,{\bf E}^{*}\bdot\frac{{\bf B}^{*}}{B_{\|}^{*}}, \label{eq:pgc_dot}
\end{eqnarray}
are expressed in terms of the guiding-center potentials
\begin{equation}
\left. \begin{array}{rcl}
e\,\Phi^{*} & \equiv & e\,\Phi \;+\; \mu\,B \\
 &  & \\
 e\,{\bf A}^{*} & \equiv & e\,{\bf A} + c\,p_{\|}\;\bhat
 \end{array} \right\},
 \label{eq:PhiA_star}
 \end{equation}
which are used to define the guiding-center electromagnetic fields
\begin{eqnarray}
{\bf E}^{*} & \equiv & -\nabla\Phi^{*} - \frac{1}{c}\pd{{\bf A}^{*}}{t} \;=\; {\bf E} - \frac{\mu}{e}\nabla B - \frac{p_{\|}}{e}\pd{\bhat}{t}, \label{eq:E_star} \\
{\bf B}^{*} & \equiv & \nabla\btimes{\bf A}^{*} \;=\; {\bf B} \;+\; \frac{c}{e}\,p_{\|}\;\nabla\btimes\bhat, \label{eq:B_star}
\end{eqnarray}
with the standard electromagnetic fields defined as
\begin{equation}
\left({\bf E},\frac{}{} {\bf B}\right) \;\equiv\; \left( -\,\nabla\Phi - \frac{1}{c}\,\pd{\bf A}{t},\; \nabla\btimes{\bf A} \right),
\label{eq:EB_PhiA}
\end{equation}
and
\begin{equation}
B_{\|}^{*} \;\equiv\; \bhat\bdot{\bf B}^{*} \;=\; B \;+\; \frac{c}{e}\;p_{\|}\;\bhat\bdot\nabla\btimes\bhat.
\label{eq:B||_star}
\end{equation}
Here, we note that, in accordance with the guiding-center approximation \cite{footnote1}, the term $B_{\|}^{*}$ does not vanish.

The guiding-center electromagnetic fields \eqref{eq:E_star}-\eqref{eq:B_star} satisfy the source-free guiding-center Maxwell equations
 \begin{eqnarray}
 \nabla\bdot{\bf B}^{*} & \equiv & \nabla\bdot{\bf B} \;=\; 0, \label{eq:div_B} \\
 \nabla\btimes{\bf E}^{*} + \frac{1}{c}\pd{{\bf B}^{*}}{t} & \equiv & \nabla\btimes{\bf E} + \frac{1}{c}\pd{{\bf B}}{t} \;=\; 0, \label{eq:curl_E}
 \end{eqnarray}
 which follow from the regular source-free Maxwell equations. Moreover, the guiding-center equations of motion \eqref{eq:Xgc_dot}-\eqref{eq:pgc_dot} satisfy the guiding-center Liouville equation
 \begin{eqnarray}
\pd{}{z^{a}}\left(B_{\|}^{*}\frac{d_{\rm gc}z^{a}}{dt} \right) & = & \nabla\btimes{\bf E}^{*}\bdot c\,\bhat \;-\; \pd{\bhat}{t}\bdot{\bf B}^{*} 
\nonumber \\
 & = & -\;\pd{B_{\|}^{*}}{t},
   \label{eq:gcLiouville}
   \end{eqnarray}
where we used Eq.~\eqref{eq:curl_E} as well as the identity $\partial{\bf B}^{*}/\partial p_{\|} \equiv (c/e)\,\nabla\btimes\bhat$. Hence, using the guiding-center Liouville equation \eqref{eq:gcLiouville}, we can write the guiding-center Vlasov equation \eqref{eq:gcV} in divergence form as
 \begin{eqnarray}
\pd{F_{\mu}}{t} & = & -\;\nabla\bdot\left( F_{\mu}\;\frac{d_{\rm gc}{\bf X}}{dt} \right) - \pd{}{p_{\|}}\left( F_{\mu}\;\frac{d_{\rm gc}p_{\|}}{dt} \right) \nonumber \\
 & \equiv & -\;\pd{}{z^{a}}\left( F_{\mu}\;\frac{d_{\rm gc}z^{a}}{dt} \right),
  \label{eq:gcV_div}
  \end{eqnarray}
where  the guiding-center Vlasov phase-space density 
\begin{equation}
F_{\mu} \;\equiv\; 2\pi\,mB_{\|}^{*}\,f_{\mu} 
\label{eq:gc_PSD}
\end{equation}
is defined in terms of the guiding-center Vlasov function $f_{\mu}$ and the guiding-center Jacobian \eqref{eq:B||_star} \cite{footnote2}.  

 \subsection{Guiding-center Maxwell equations}
 
 The source-free Maxwell equations \eqref{eq:div_B}-\eqref{eq:curl_E} are complemented by the guiding-center Maxwell equations
 \begin{eqnarray}
 \nabla\bdot{\bf E} & = & 4\pi\;\varrho_{\rm gc}, \label{eq:div_E} \\
 \nabla\btimes{\bf B} - \frac{1}{c}\pd{\bf E}{t} & = & \frac{4\pi}{c} \left( {\bf J}_{\rm gc} \;+\frac{}{} c\,\nabla\btimes{\bf M}_{\rm gc} \right), \label{eq:curl_B}
 \end{eqnarray}
  where the guiding-center charge and current densities 
 \begin{equation}
 (\varrho_{\rm gc},\; {\bf J}_{\rm gc}) \;\equiv\; \int  \left( e,\; e\,\frac{d_{\rm gc}{\bf X}}{dt} \right)\;F_{\mu}\, dp_{\|}\,d\mu
 \label{eq:rhoJ_gc}
 \end{equation}
 are expressed as moments of the guiding-center Vlasov phase-space density $F_{\mu}$, and summation over particle species is assumed (wherever appropriate) throughout the text. We note that the guiding-center Maxwell equations \eqref{eq:div_E}-\eqref{eq:curl_B} imply that the guiding-center charge and current densities \eqref{eq:rhoJ_gc} satisfy the guiding-center charge conservation law $\partial\varrho_{\rm gc}/\partial t + \nabla\bdot{\bf J}_{\rm gc} = 0$.
   
The guiding-center magnetization current density $c\,\nabla\btimes{\bf M}_{\rm gc}$ in Eq.~\eqref{eq:curl_B}  is defined in terms of the guiding-center magnetization \cite{Brizard_2008,Brizard_2009}
 \begin{equation}
 {\bf M}_{\rm gc} \equiv \int  \left( \vb{\mu}_{\rm gc} + \vb{\pi}_{\rm gc}\btimes \frac{p_{\|}\,\bhat}{mc} \right)\;F_{\mu}\, dp_{\|}\,d\mu,
 \label{eq:Mgc}
 \end{equation}
 which is expressed as the sum of an intrinsic contribution due to the magnetic-dipole moment $\vb{\mu}_{\rm gc}$ and a moving electric-dipole contribution due to the guiding-center electric-dipole moment $\vb{\pi}_{\rm gc}$, where \cite{Brizard_2013,Tronko_Brizard_2015}
 \begin{eqnarray}
\vb{\mu}_{\rm gc} & \equiv & -\;\mu\,\bhat, \label{eq:mu_gc} \\
\vb{\pi}_{\rm gc} & \equiv & \frac{e\bhat}{\Omega}\btimes \frac{d_{\rm gc}{\bf X}}{dt}. \label{eq:pi_gc}
  \end{eqnarray}
As mentioned above, we are using a formulation of guiding-center Vlasov-Maxwell theory where the guiding-center polarization current is absent and the guiding-center polarization enters only as a contribution to the guiding-center magnetization \eqref{eq:Mgc}.
  
  \subsection{Organization}
  
  We now show that the guiding-center Vlasov-Maxwell equations \eqref{eq:gcV} and \eqref{eq:div_E}-\eqref{eq:curl_B} possess Lagrange, Euler, and Euler-Poincar\'{e} variational formulations. The reader is invited to consult Newcomb's pioneering work \cite{Newcomb_1962}, as well as review papers by Morrison \cite{Morrison_1998,Morrison_2005}, to review the distinction between the Lagrangian and Eulerian formulations of classical mechanics as applied to plasma physics. 
  
The remainder of this paper is organized as follows. In Sec.~\ref{sec:Lagrangian}, the Lagrange variational principle for the guiding-center Vlasov-Maxwell equations is presented, based on the standard Low-Lagrangian formalism \cite{Low_1958,Kaufman_Holm_1984, Similon_1985,Ye_Kaufman_1992, Ye_Morrison_1992} involving a Hamilton-Maxwell action functional. The variational fields in the guiding-center Hamilton-Maxwell action functional are the guiding-center phase-space coordinates ${\bf z}$ and their ``velocities'' $d{\bf z}/dt$, which appear in the guiding-center Hamilton action functional, with the electromagnetic fields $({\bf E},{\bf B})$ appearing in both the Hamilton and Maxwell action functionals.  
    
  In Sec.~\ref{sec:Eulerian}, the Euler variational principle for the guiding-center Vlasov-Maxwell equations is presented, based on the standard Eulerian formalism \cite{Brizard_2000} involving a Vlasov-Maxwell action functional that depends on the variational fields $({\cal F}_{\mu},{\bf E},{\bf B})$, where 
  ${\cal F}_{\mu}$ denotes the extended Vlasov phase-space distribution. While the intrinsic magnetic-dipole term \eqref{eq:mu_gc} still appears through the variation of the Hamiltonian with respect to the magnetic field, the moving electric-dipole term \eqref{eq:pi_gc} appears through the variation of the Vlasov distribution. 
    
  In Sec.~\ref{sec:Noether}, we derive the energy-momentum conservation laws for the guiding-center Vlasov-Maxwell equations based on the guiding-center Noether equation derived from the Eulerian variational principle presented in Sec.~\ref{sec:Eulerian}. Here, we show how important the moving electric dipole-moment term \eqref{eq:pi_gc} contributes to the conservation of energy and momentum.
  
  In Sec.~\ref{sec:EP}, the Euler-Poincar\'{e} variational formulation of the guiding-center Vlasov-Maxwell equations is presented, along with its Kelvin-Noether conservation laws recovering the symplectic features of the guiding-center dynamics.
  
  \section{\label{sec:Lagrangian}Lagrange Variational Principle}
 
 The Lagrangian formulation of guiding-center dynamics is expressed in terms of a Lagrangian path 
 \begin{equation}
 z^{a}(t;{\bf z}_{0},\mu) \;\equiv\; \left({\bf X}(t;{\bf z}_{0},\mu),\frac{}{} p_{\|}(t;{\bf z}_{0},\mu)\right)
 \label{eq:Lag_z}
 \end{equation}
 in four-dimensional reduced guiding-center phase space, which is uniquely defined by the labels $({\bf z}_{0},\mu)$ represented by the initial phase-space position ${\bf z}_{0} =
  ({\bf X}_{0},p_{\|0}) \equiv  {\bf z}(0;{\bf z}_{0},\mu)$ and the magnetic-moment invariant $\mu$. A Lagrangian ``velocity'' $\dot{\bf z} \equiv \partial{\bf z}/\partial t$ is also defined along each Lagrangian path ${\bf z}(t;{\bf z}_{0},\mu)$, which can then be used to form an eight-dimensional guiding-center phase space, with coordinates $({\bf z},\dot{\bf z})$ labeled by $({\bf z}_{0}, \mu)$.
 
The phase-space Lagrangian variational principle for the guiding-center Vlasov-Maxwell equations \eqref{eq:gcV} and \eqref{eq:div_E}-\eqref{eq:curl_B} is based on the guiding-center Hamilton-Maxwell functional \cite{Similon_1985,Ye_Kaufman_1992}
  \begin{eqnarray}
 {\cal A}_{\rm gc}^{\rm L} & = &  \int \;L_{\rm gc}({\bf z},\dot{{\bf z}}; \Phi, {\bf A}, {\bf B})\; dt,
   \nonumber \\
   &  &+\; \int  \frac{d^{3}x\;dt}{8\pi} \left( |{\bf E}|^{2} \;-\frac{}{} |{\bf B}|^{2} \right)
  \label{eq:Agc_Lag}   
  \end{eqnarray}
   where the fields $(\mathbf{E},\mathbf{B})$ are defined in terms of the potentials $(\Phi,{\bf A})$ according to Eq.~\eqref{eq:EB_PhiA}. Also, the guiding-center Lagrangian
  \begin{equation}
  L_{\rm gc} \;\equiv\; \int \;\Lambda_{\rm gc}({\bf z},\dot{{\bf z}}; \Phi, {\bf A}, {\bf B})\;F_{0}({\bf z}_{0},\mu)\;d^{4}z_{0}\,d\mu 
  \label{eq:Lgc_Lag}
  \end{equation}
 is expressed in terms of an average over the initial conditions $({\bf z}_{0},\mu)$, with an initial phase-space density $F_{0}({\bf z}_{0},\mu)$, of the guiding-center single-particle phase-space Lagrangian \cite{Littlejohn}
\begin{equation}
\Lambda_{\rm gc} \;\equiv\; \frac{e}{c}\,{\bf A}^{*}\bdot\dot{\bf X} \;-\; \left( \frac{p_{\|}^{2}}{2m} \;+\; e\,\Phi^{*} \right).
\label{eq:Lambda_gc}
\end{equation}
The guiding-center Lagrangian \eqref{eq:Lambda_gc} also depends on the electromagnetic potentials $(\Phi,{\bf A})$, which yields the guiding-center charge and current densities
\eqref{eq:rhoJ_gc}, as well as the magnetic field ${\bf B}$ through the definitions \eqref{eq:PhiA_star} for $(\Phi^{*},{\bf A}^{*})$, which yields the guiding-center dipole contributions
\eqref{eq:mu_gc}-\eqref{eq:pi_gc} to the guiding-center magnetization \eqref{eq:Mgc}. 
 
 \subsection{Guiding-center Vlasov equation}
 
 We first derive the guiding-center equations of motion \eqref{eq:Xgc_dot}-\eqref{eq:pgc_dot} for the guiding-center position ${\bf X}$ and the guiding-center parallel momentum $p_{\|}$ from the Lagrangian \eqref{eq:Lambda_gc} as Euler-Lagrange equations. These guiding-center equations of motion are then used as the characteristics of the guiding-center Vlasov equation \eqref{eq:gcV} for the guiding-center Vlasov function $f_{\mu}({\bf z}(t;{\bf z}_{0},\mu),t)$.
  
From Eq.~\eqref{eq:Lambda_gc}, we derive the guiding-center Euler-Lagrange equations
\begin{equation}
\frac{d}{dt}\left(\pd{\Lambda_{\rm gc}}{\dot{z}^{a}}\right) \;=\; \pd{\Lambda_{\rm gc}}{z^{a}},
\label{eq:EL_za}
\end{equation}
from which we derive expressions $d_{\rm gc}z^{a}/dt$ for the characteristics of the guiding-center Vlasov equation \eqref{eq:gcV}.  The Euler-Lagrange equation associated with variations $\delta{\bf X}$ is derived from Eq.~\eqref{eq:Lambda_gc} as
 \begin{eqnarray}
 0 & = & \nabla \Lambda_{\rm gc} \;-\; \frac{d}{dt}\left(\pd{\Lambda_{\rm gc}}{\dot{\bf X}}\right) \nonumber \\
  & = & e\;\left( \frac{1}{c}\,\nabla{\bf A}^{*}\bdot\dot{\bf X} \;-\; \nabla\Phi^{*} \right) \;-\; \frac{e}{c}\,\frac{d{\bf A}^{*}}{dt} \nonumber \\
  & \equiv & e\,{\bf E}^{*} \;+\; \frac{e}{c}\,\dot{\bf X}\btimes{\bf B}^{*} \;-\; \dot{p}_{\|}\;\bhat,
  \label{eq:EL_X}
  \end{eqnarray} 
  where $\nabla$ is evaluated at constant $p_{\|}$ and $(\dot{\bf X},\dot{p}_{\|})$. 
  
  Next, since the guiding-center Lagrangian \eqref{eq:Lambda_gc} is independent of $\dot{p}_{\|}$, the Euler-Lagrange equation associated with variations $\delta p_{\|}$ is simply expressed as the Lagrangian constraint equation
 \begin{equation}
 0 \;\equiv\; \pd{\Lambda_{\rm gc}}{p_{\|}} \;=\; \bhat\bdot\dot{\bf X} \;-\; \frac{p_{\|}}{m}.
 \label{eq:EL_p}
 \end{equation}
By taking the cross product of the Euler-Lagrange equation \eqref{eq:EL_X} with $\bhat$, and using the constraint \eqref{eq:EL_p}, we obtain the expression \eqref{eq:Xgc_dot} for $d_{\rm gc}{\bf X}/dt$. By taking the dot product of Eq.~\eqref{eq:EL_X} with ${\bf B}^{*}$, on the other hand, we obtain the expression \eqref{eq:pgc_dot} for $d_{\rm gc}p_{\|}/dt$.
 
 \subsection{Guiding-center Maxwell equations}
 
 The variation of the guiding-center Hamilton-Maxwell action functional \eqref{eq:Agc_Lag} with respect to the electromagnetic-field variations yields
 \begin{eqnarray}
 \delta{\cal A}_{\rm gc}^{\rm L} & = & \int  d^{3}x\;dt \left[ \frac{1}{4\pi}\left( \delta{\bf E}\bdot{\bf E} \;-\frac{}{} \delta{\bf B}\bdot{\bf H} \right) \right. 
 \nonumber \\
  &  &\left.\hspace*{0.3in}+\; \left( \frac{1}{c}\delta{\bf A}\bdot {\bf J}_{\rm gc} - \delta\Phi\,\varrho_{\rm gc} \right) \right],
  \label{eq:delta_Agc_L}
 \end{eqnarray}
 where we introduced the ``macroscopic'' magnetic field
 \begin{equation}
 {\bf H} \;\equiv\; {\bf B} \;-\; 4\pi\,{\bf M}_{\rm gc},
 \label{eq:Hgc_field}
 \end{equation} 
 and the guiding-center charge and current densities
  \begin{equation}
 (\varrho_{\rm gc},\; {\bf J}_{\rm gc}) \;\equiv\; \int _{0} \delta^{3}({\bf x} - {\bf X})\;\left( e,\; e\,\frac{d_{\rm gc}{\bf X}}{dt} \right),
 \label{eq:rhoJ_gcL}
 \end{equation}
where $\int _{0} (\cdots) \equiv \int  (\cdots)\,F_{0}\,d^{4}z_{0}\,d\mu$. The guiding-center magnetization \eqref{eq:Mgc}, on the other hand, is defined in terms of the guiding-center Lagrangian \eqref{eq:Lambda_gc} as
\begin{equation}
 {\bf M}_{\rm gc} \equiv \int _{0} \delta^{3}({\bf x} - {\bf X})\;\pd{\Lambda_{\rm gc}}{\bf B},
 \label{eq:Mgc_L}
 \end{equation}
where the Lagrangian magnetic derivative
\begin{eqnarray}
\pd{\Lambda_{\rm gc}}{\bf B} & = & \frac{e}{c}\,\pd{{\bf A}^{*}}{\bf B}\bdot\dot{\bf X} - e\;\pd{\Phi^{*}}{\bf B} = p_{\|}\,\pd{\bhat}{\bf B}\bdot
\frac{d_{\rm gc}{\bf X}}{dt} - \mu\,\pd{B}{\bf B} \nonumber \\
 & = & \left(\bhat\btimes\frac{d_{\rm gc}{\bf X}}{dt}\right)\btimes\frac{p_{\|}\,\bhat}{B} \;-\; \mu\;\bhat \nonumber \\
  & \equiv & \vb{\pi}_{\rm gc}\btimes\frac{p_{\|}\,\bhat}{mc} \;+\; \vb{\mu}_{\rm gc}
 \label{eq:Lgc_B}
\end{eqnarray}
yields the guiding-center intrinsic magnetic and moving-electric dipole contributions \eqref{eq:mu_gc}-\eqref{eq:pi_gc}.

We now need to express the variation \eqref{eq:delta_Agc_L} solely in terms of the variations $(\delta\Phi, \delta{\bf A})$, which amounts to using the electromagnetic-variation constraints
 \begin{equation}
 \delta{\bf E} \equiv -\,\nabla\delta\Phi - \frac{1}{c}\pd{\delta{\bf A}}{t} \;\;{\rm and}\;\; \delta{\bf B} \equiv \nabla\btimes\delta{\bf A},
 \label{eq:delta_EB}
 \end{equation}
thereby guaranteeing that the source-free Maxwell equations \eqref{eq:div_B}-\eqref{eq:curl_E} are preserved
 \[ \nabla\bdot\delta{\bf B} \;=\; 0 \;=\; \frac{1}{c}\,\pd{\delta{\bf B}}{t} \;+\; \nabla\btimes\delta{\bf E}. \]
These equations are equivalent to stating that the variation of the Faraday tensor $\delta{\sf F}$ is closed $\exd\delta{\sf F} = 0$, which implies that it is an exact space-time two-form $\delta{\sf F} \equiv \exd\delta{\sf A}$, where the variation of the standard electromagnetic-potential one-form is $\delta{\sf A} \equiv \delta{\bf A}\bdot\exd{\bf x} - \delta\Phi\,c\,\exd t$.
 
 If we now substitute Eq.~\eqref{eq:delta_EB} into the action-functional variation \eqref{eq:delta_Agc_L}, we obtain
 \begin{eqnarray}
 \delta{\cal A}_{\rm gc}^{\rm L} & = & \left. \int  \frac{d^{3}x\,dt}{4\pi} \right[ \delta\Phi\frac{}{} \left( \nabla\bdot{\bf E} \;-\frac{}{} 4\pi\,\varrho_{\rm gc} \right) 
\\
  &  &\hspace*{0.5in}\left.+\;\delta{\bf A}\bdot\left( \frac{4\pi}{c}\,{\bf J}_{\rm gc} + \frac{1}{c}\pd{\bf E}{t} - \nabla\btimes{\bf H} \right) \right],
 \nonumber 
  \end{eqnarray}
 where an exact space-time derivative, which is not shown here, vanishes upon integration. Stationarity of the guiding-center Hamilton-Maxwell action functional with respect to arbitrary variations $(\delta\Phi, \delta{\bf A})$ yields the guiding-center Maxwell equations \eqref{eq:div_E}-\eqref{eq:curl_B}.
  
\section{\label{sec:Eulerian}Euler Variational Principle}  

The Eulerian representation of guiding-center particle dynamics replaces the Lagrangian guiding-center coordinates \eqref{eq:Lag_z} with fixed Eulerian guiding-center coordinates 
$\zeta^{a} \equiv (\vb{\sf X},{\sf p}_{\|})$, with the magnetic moment $\mu$ once again acting as a label. Since a unique Lagrangian path ${\bf z}(t;{\bf z}_{0},\mu)$, labeled by 
$({\bf z}_{0},\mu)$, passes through each Eulerian point $\vb{\zeta}$ in guiding-center phase space, there is, thus, a one-to-one mapping between the Lagrangian and Eulerian coordinates.

The Eulerian variational principle for the guiding-center Vlasov-Maxwell equations is based on the guiding-center Vlasov-Maxwell action functional
\cite{Brizard_2000}
\begin{eqnarray}
{\cal A}_{\rm gc}^{\rm E} & = &  \int  \frac{d^{3}x\;dt}{8\pi} \left( |{\bf E}|^{2} \;-\frac{}{} |{\bf B}|^{2} \right)
\nonumber \\
   &  &+\; \int  {\cal F}_{\mu}\,{\cal H}\;d^{6}Z\,d\mu,
  \label{eq:Agc_Eul} 
   \end{eqnarray}
   where the Eulerian fields ${\cal F}_{\mu}$ and ${\cal H}$ depend on the phase-space coordinates $(Z^{\alpha}; \mu) \equiv (\vb{\sf X}, {\sf p}_{\|}, w, t; \mu)$ in extended
   $(6+1)$-dimensional guiding-center phase space, which include the guiding-center Eulerian coordinates $\zeta^{a} \equiv (\vb{\sf X},{\sf p}_{\|})$ and 
   the canonically-conjugate energy-time coordinates $(w,t)$.  Here, in contrast to the Lagrangian representation used in Sec.~\ref{sec:Lagrangian}, the Eulerian coordinates 
   $\vb{\zeta}$ used here are independent of time.
     
  The extended guiding-center Hamiltonian in Eq.~\eqref{eq:Agc_Eul} 
  \begin{equation}
  {\cal H} \;\equiv\; \left( e\,\Phi^{*} + \frac{{\sf p}_{\|}^{2}}{2m}\right) \;-\; w \;=\; H_{\rm gc} \;-\; w,
  \label{eq:Hgc_ext}
  \end{equation}
  is expressed in terms of the guiding-center Hamiltonian $H_{\rm gc}$ and the guiding-center energy coordinate $w$. The extended guiding-center Vlasov phase-space density
  \begin{equation}
  {\cal F}_{\mu} \;\equiv\; F_{\mu}(\vb{\sf X}, {\sf p}_{\|}, t)\; \delta(w - H_{\rm gc}),
  \label{eq:Fgc_ext}
  \end{equation}
where $F_{\mu}$ is defined in Eq.~\eqref{eq:gc_PSD}, ensures that the physical guiding-center motion in extended phase space takes places on the guiding-center energy surface 
${\cal H} \equiv 0$ so that Eq.~\eqref{eq:Hgc_ext} yields $w = H_{\rm gc}$. The extended guiding-center Vlasov phase-space density \eqref{eq:Fgc_ext} yields the following integral identity
\begin{equation}
\int {\cal F}_{\mu}\,{\cal H}\;{\cal G} \,dw \;=\; \int F_{\mu}\;\delta({\cal H})\;{\cal H}\;{\cal G}\;dw \;\equiv\; 0,
\label{eq:FH_id}
\end{equation}
where ${\cal G}$ denotes an arbitrary function.

The variation of the guiding-center Vlasov-Maxwell action functional \eqref{eq:Agc_Eul} yields
\begin{eqnarray}
\delta{\cal A}_{\rm gc}^{\rm E} & = & \int  \frac{d^{3}x\;dt}{4\pi} \left( {\bf E}\bdot\delta{\bf E} \;-\frac{}{} {\bf B}\bdot\delta{\bf B} \right)
\label{eq:deltaAgc_E} \\
   &  &+\; \int  \left( \delta{\cal F}_{\mu}\,{\cal H} \;+\frac{}{} {\cal F}_{\mu}\;\delta{\cal H} \right)\;d^{6}Z\,d\mu,
\nonumber
\end{eqnarray}
where the Eulerian variations
\begin{eqnarray}
\delta{\cal H} & \equiv & e\,\delta\Phi^{*}, \label{eq:delta_Phi} \\
\delta{\cal F}_{\mu} & \equiv & \frac{e}{c}\,\delta{\bf A}^{*}\bdot\left(B_{\|}^{*}\left\{ \vb{\sf X},\;  {\cal F}_{\mu}/B_{\|}^{*}\right\}_{\rm gc} \right) \;+\; \delta B_{\|}^{*}\; 
{\cal F}_{\mu}/B_{\|}^{*} \nonumber \\
 &  &+\; B_{\|}^{*}\;\left\{ \delta{\cal S},\; {\cal F}_{\mu}/B_{\|}^{*}\right\}_{\rm gc},
\label{eq:delta_F}
\end{eqnarray}
are expressed in terms of
\begin{eqnarray}
e\,\delta\Phi^{*} & = & e\,\delta\Phi \;+\; \mu\,\bhat\bdot\delta{\bf B}, \label{eq:delta_Phi} \\
\frac{e}{c}\,\delta{\bf A}^{*} & = & \frac{e}{c}\,\delta{\bf A} \;+\; {\sf p}_{\|}\,\delta{\bf B}\bdot\pd{\bhat}{\bf B},
\label{eq:delta_A}
\end{eqnarray}
with 
\begin{equation}
\delta B_{\|}^{*} \;\equiv\; \delta{\bf B}^{*}\bdot\bhat \;+\; \left(\delta{\bf B}\bdot\pd{\bhat}{\bf B}\right)\bdot{\bf B}^{*}. 
\label{eq:delta_Bstar}
\end{equation}
In order to make these expressions explicit, however, we need to derive the guiding-center Poisson bracket $\{\;,\;\}_{\rm gc}$ in extended guiding-center phase space. 

The first two terms on the right side of Eq.~\eqref{eq:delta_F} appear as a result of the noncanonical nature of the guiding-center formalism. In addition, we will show below [see 
Eq.~\eqref{eq:delta_BF}] that the second term on the right side of Eq.~\eqref{eq:delta_F} is needed to ensure that the particle-number conservation law
\begin{equation}
 \int  \delta{\cal F}_{\mu}\;d^{6}Z\,d\mu \;\equiv\; 0
 \label{eq:delta_F_constraint}
 \end{equation}
is satisfied. We note that the last term on the right side of Eq.~\eqref{eq:delta_F} represents the canonical component of the variation of the extended phase-space density 
${\cal F}_{\mu}$, which independently satisfies the constraint \eqref{eq:delta_F_constraint}. 

\subsection{Guiding-center Poisson bracket}

The guiding-center Poisson bracket $\{\;,\;\}_{\rm gc}$ used in Eq.~\eqref{eq:delta_F} is derived from the extended guiding-center one-form \cite{footnote3} 
\begin{equation}
\Gamma_{\rm gc} \;\equiv\; \frac{e}{c}\,{\bf A}^{*}\bdot\exd\vb{\sf X} \;-\; w\;\exd t.
\label{eq:Gamma_gc}
\end{equation}
First, we derive the extended guiding-center Lagrange two-form
\begin{eqnarray}
\vb{\omega}_{\rm gc} \equiv \exd\Gamma_{\rm gc} & = & \frac{e}{c} \left( \exd {\sf X}^{i}\,\partial_{i}A_{j}^{*} \;+\frac{}{} \exd t\,\partial_{t}A_{j}^{*}\right)\wedge 
\exd {\sf X}^{j}
\nonumber \\
 &  &+\; \exd {\sf p}_{\|}\wedge \bhat\bdot\exd\vb{\sf X} \;-\; \exd w\wedge \exd t.
 \label{eq:omega_gc}
 \end{eqnarray}
 Since the determinant $(e B_{\|}^{*}/c)^{2} \neq 0$ of the $6\times 6$ Lagrange matrix constructed from the Lagrange two-form \eqref{eq:omega_gc} does not vanish, the Lagrange matrix can be inverted to obtain the extended guiding-center Poisson matrix $J_{\rm gc}^{\alpha\beta}$, from which we construct the extended guiding-center Poisson bracket
\[ \{ {\cal F},\; {\cal G}\}_{\rm gc} \;\equiv\; \pd{\cal F}{Z^{\alpha}}\;J_{\rm gc}^{\alpha\beta}\;\pd{\cal G}{Z^{\beta}}. \]
The extended guiding-center Poisson bracket is, thus, constructed as
 \begin{eqnarray}
 \{ {\cal F},\; {\cal G}\}_{\rm gc}  & = & \frac{{\bf B}^{*}}{B_{\|}^{*}}\bdot\left(\nabla^{*}{\cal F}\;\pd{\cal G}{{\sf p}_{\|}} \;-\; \pd{\cal F}{{\sf p}_{\|}}\;\nabla^{*}{\cal G} \right) \nonumber \\
   &  &-\; \frac{c\,\bhat}{eB_{\|}^{*}}\bdot\nabla^{*}{\cal F}\btimes\nabla^{*}{\cal G} \nonumber \\
   &  &+\; \left( \pd{\cal F}{w}\;\pd{\cal G}{t} \;-\; \pd{\cal F}{t}\;\pd{\cal G}{w} \right),
   \label{eq:gcPB_ext}
   \end{eqnarray}
 where we have defined the effective gradient
 \begin{equation}
 \nabla^{*}{\cal F} \;\equiv\; \nabla{\cal F} \;-\; \frac{e}{c}\,\pd{{\bf A}^{*}}{t}\;\pd{\cal F}{w},
 \label{eq:grad_star}
 \end{equation}
 so that $\nabla^{*}{\cal H} = e\,\nabla\Phi^{*} + (e/c)\,\partial{\bf A}^{*}/\partial t \equiv -\,e\,{\bf E}^{*}$. The fundamental Poisson brackets $\{Z^{\alpha},\; {\cal G}\}_{\rm gc}$ are  \begin{eqnarray}
 \{ \vb{\sf X},\; {\cal G}\}_{\rm gc} & = & \frac{{\bf B}^{*}}{B_{\|}^{*}}\,\pd{\cal G}{{\sf p}_{\|}} \;+\; \frac{c\,\bhat}{eB_{\|}^{*}}\btimes\nabla^{*}{\cal G}, \label{eq:PB_X} \\
 \{ {\sf p}_{\|},\; {\cal G}\}_{\rm gc} & = & -\;\frac{{\bf B}^{*}}{B_{\|}^{*}}\bdot\nabla^{*}{\cal G}, \label{eq:PB_p} \\
 \{ w,\; {\cal G}\}_{\rm gc} & = & \pd{\cal G}{t} \;-\; \frac{e}{c}\pd{{\bf A}^{*}}{t}\bdot\{\vb{\sf X},\; {\cal G}\}_{\rm gc}, \label{eq:PB_w} \\
 \{ t,\; {\cal G}\}_{\rm gc} & = & -\;\pd{\cal G}{w}, \label{eq:PB_t}
 \end{eqnarray}
where the function ${\cal G}$ is arbitrary. 

We note that the extended guiding-center Poisson bracket \eqref{eq:gcPB_ext} satisfies the standard properties of a Poisson bracket: (i) antisymmetry property, 
$\{{\cal F},\;{\cal G}\}_{\rm gc} = -\,\{{\cal G},\; {\cal F}\}_{\rm gc}$; (ii) Leibniz property, $\{{\cal F},\;({\cal G}\,{\cal K})\}_{\rm gc} = (\{{\cal F},\;{\cal G}\}_{\rm gc})
{\cal K} + {\cal G}(\{{\cal F},\;{\cal K}\}_{\rm gc})$; and (iii) Jacobi identity,
\begin{eqnarray}
0 & = & \left\{ {\cal F},\frac{}{} \{{\cal G},\; {\cal K}\}_{\rm gc}\right\}_{\rm gc} \;+\; \left\{ {\cal G},\frac{}{} \{{\cal K},\; {\cal F}\}_{\rm gc}\right\}_{\rm gc} \nonumber \\
 &  &+\; \left\{ {\cal K},\frac{}{} \{{\cal F},\; {\cal G}\}_{\rm gc}\right\}_{\rm gc}.
\end{eqnarray}
This identity follows from the fact that the Lagrange two-form \eqref{eq:omega_gc} on extended guiding-center phase space is closed: $\exd\vb{\omega}_{\rm gc}
= \exd^{2}\Gamma_{\rm gc} \equiv 0$ since it is exact (assuming that $\nabla\bdot{\bf B}^{*} \equiv 0$ is satisfied).

 With the extended guiding-center Poisson bracket \eqref{eq:gcPB_ext}, the guiding-center equations of motion 
 \eqref{eq:Xgc_dot}-\eqref{eq:pgc_dot} can be expressed in Hamiltonian form as
 \begin{eqnarray}
 \frac{d_{\rm gc}{\bf X}}{dt} & \equiv & \{ \vb{\sf X},\; {\cal H}\}_{\rm gc} \;=\; \frac{{\sf p}_{\|}}{m}\,\frac{{\bf B}^{*}}{B_{\|}^{*}} \;+\; {\bf E}^{*}\btimes
 \frac{c\,\bhat}{B_{\|}^{*}}, \label{eq:Xgc_H} \\
 \frac{d_{\rm gc}p_{\|}}{dt} & \equiv & \{ {\sf p}_{\|},\; {\cal H}\}_{\rm gc} \;=\; {\bf E}^{*}\bdot\frac{{\bf B}^{*}}{B_{\|}^{*}}.
 \label{eq:pgc_H}
 \end{eqnarray}
Hence, the ``Lagrangian'' velocities \eqref{eq:Xgc_H}-\eqref{eq:pgc_H} are represented here as ``Eulerian'' functions. The guiding-center equations of motion for
$(w,t)$, on the other hand, are expressed in Hamiltonian form as
 \begin{eqnarray}
 \frac{d_{\rm gc}t}{dt} & \equiv & \{ t,\; {\cal H}\}_{\rm gc} \;=\; -\;\pd{\cal H}{w} \;=\; 1, \label{eq:tgc_H} \\ 
 \frac{d_{\rm gc}w}{dt} & \equiv & \{ w,\; {\cal H}\}_{\rm gc} \;=\; e\,\pd{\Phi^{*}}{t} - \frac{e}{c}\,\pd{{\bf A}^{*}}{t}\bdot\frac{d_{\rm gc}{\bf X}}{dt},
 \label{eq:wgc_H}
 \end{eqnarray}
 where we used $\nabla^{*}w = -\,(e/c)\,\partial{\bf A}^{*}/\partial t$. These equations can be used to show that the total energy ${\cal H} \equiv H_{\rm gc} - w$ is conserved: 
 $d_{\rm gc}{\cal H}/dt = 0$.
 
The extended guiding-center Poisson bracket \eqref{eq:gcPB_ext} can also be expressed in phase-space divergence form
 \begin{equation}
\{ {\cal F},\; {\cal G}\}_{\rm gc} \;\equiv\; \frac{1}{B_{\|}^{*}}\pd{}{Z^{\alpha}}\left(B_{\|}^{*}\,{\cal F}\frac{}{} \{ Z^{\alpha},\; {\cal G}\}_{\rm gc} \right),
 \label{eq:gcPB_div}
 \end{equation}
 which yields the identity
 \begin{equation}
 \int {\cal K} \{ {\cal F},\; {\cal G}\}_{\rm gc}B_{\|}^{*}\,d^{6}Z = -\,\int {\cal F}\{ {\cal K},\; {\cal G}\}_{\rm gc}B_{\|}^{*}\,d^{6}Z,
 \label{eq:gcPB_id}
 \end{equation}
 where ${\cal F}$, ${\cal G}$, and ${\cal K}$ are arbitrary functions.

The divergence form of the extended guiding-center Poisson bracket \eqref{eq:gcPB_div} can thus be used to write the useful expression
 \begin{equation}
B_{\|}^{*} \left\{ {\cal F}_{\mu}/B_{\|}^{*}, {\cal H}\right\}_{\rm gc} \;\equiv\; \pd{}{Z^{\alpha}}\left({\cal F}_{\mu}\;\frac{d_{\rm gc}Z^{\alpha}}{dt}\right),
\label{eq:F_star}
 \end{equation}
where $d_{\rm gc}Z^{\alpha}/dt$ are given by Eqs.~\eqref{eq:Xgc_H}-\eqref{eq:wgc_H}. Next, when Eq.~\eqref{eq:F_star} is integrated upon $w$, we obtain the phase-space divergence form of the guiding-center Vlasov equation \eqref{eq:gcV_div}:
 \begin{equation}
\int  B_{\|}^{*}\{ {\cal F}_{\mu}/B_{\|}^{*}, {\cal H}\}_{\rm gc}\,dw = \pd{F_{\mu}}{t} + \pd{}{z^{a}}\left( F_{\mu}\frac{d_{\rm gc}z^{a}}{dt} \right) \equiv 0,
 \label{eq:ext_PB_HF}
 \end{equation}
 where we used the definition \eqref{eq:Fgc_ext} as well as the Lagrangian velocities \eqref{eq:Xgc_H}-\eqref{eq:pgc_H}.
 
Lastly, we note that the variation \eqref{eq:delta_F} satisfies the particle-number constraint \eqref{eq:delta_F_constraint}, since, using Eqs.~\eqref{eq:PB_X} and \eqref{eq:gcPB_div}, we find
 \begin{equation}
 \delta{\cal F}_{\mu}  \;\equiv\; \pd{}{Z^{\alpha}}\left( {\cal F}_{\mu}\frac{}{}\delta Z^{\alpha}\right), 
 \label{eq:delta_BF}
 \end{equation} 
 where 
 \[ \delta Z^{\alpha} \;\equiv\; \left\{\delta{\cal S}, Z^{\alpha}\right\}_{\rm gc} + \frac{e}{c}\delta{\bf A}^{*}\bdot\left\{\vb{\sf X},Z^{\alpha}\right\}_{\rm gc} \]
and
\[ \pd{}{Z^{\alpha}}\left( \frac{e}{c}\,\delta{\bf A}^{*}\bdot B_{\|}^{*}\left\{\vb{\sf X},\frac{}{} Z^{\alpha}\right\}_{\rm gc}\right) \;\equiv\; \delta B_{\|}^{*}, \]
where we used Eq.~\eqref{eq:delta_Bstar}. Hence,  $\delta{\cal F}_{\mu}$ is expressed as an exact phase-space divergence whose phase-space integral vanishes, which satisfies the particle-number constraint \eqref{eq:delta_F_constraint}.
  
\subsection{Guiding-center Lagrangian density}

From the guiding-center Euler action functional \eqref{eq:deltaAgc_E}, we extract the Euler guiding-center Lagrangian density
\begin{equation}
{\cal L}_{\rm gc} \;\equiv\; \frac{1}{8\pi} \left( |{\bf E}|^{2} \;-\frac{}{} |{\bf B}|^{2} \right) \;-\; \int  {\cal F}_{\mu}\,{\cal H}\;d^{3}P,
\label{eq:Lag_gc}
\end{equation}
where any spatial dependence in the guiding-center Vlasov part is now evaluated at the space-time field point $({\bf x},t)$ and we use the notation
$d^{3}P \equiv d{\sf p}_{\|}\,d\mu\,dw$. The variation of the guiding-center Lagrangian density \eqref{eq:Lag_gc} yields the expression
\begin{eqnarray}
\delta{\cal L}_{\rm gc} & = & \frac{1}{4\pi} \left( \delta{\bf E}\bdot{\bf E} \;-\frac{}{} \delta{\bf B}\bdot{\bf B} \right) \nonumber \\
 &  &-\; \int   \left( \delta{\cal F}_{\mu}\;{\cal H} \;+\frac{}{} {\cal F}_{\mu}\;e\,\delta\Phi^{*} \right)\,d^{3}P
 \label{eq:delta_Lgc_def} \\
  & = & \frac{1}{4\pi} \left( \delta{\bf E}\bdot{\bf E} -\frac{}{} \delta{\bf B}\bdot{\bf B} \right) \nonumber \\
    &  &-\; \int  \left( e\,\delta\Phi^{*} - \frac{e}{c}\,\delta{\bf A}^{*}\bdot\frac{d_{\rm gc}{\bf X}}{dt}\right) {\cal F}_{\mu} \,d^{3}P \nonumber \\
   &  &-\; \int  B_{\|}^{*}\;\delta{\cal S}\, \left\{{\cal F}_{\mu}/B_{\|}^{*},\frac{}{} {\cal H} \right\}_{\rm gc} \,d^{3}P \nonumber \\
    &  &-\; \int  B_{\|}^{*}\;\left\{ \delta{\cal S}\,{\cal H},\frac{}{} {\cal F}_{\mu}/B_{\|}^{*} \right\}_{\rm gc}\,d^{3}P,
   \nonumber
 \end{eqnarray}
where the guiding-center velocity $d_{\rm gc}{\bf X}/dt$ is defined by Eq.~\eqref{eq:Xgc_H} and we used the expression
\begin{eqnarray}
\delta{\cal F}_{\mu}\,{\cal H} & = & -\;{\cal F}_{\mu}\;\frac{e}{c}\,\delta{\bf A}^{*}\bdot\frac{d_{\rm gc}{\bf X}}{dt} \;+\; B_{\|}^{*}\,\delta{\cal S}\;\left\{ {\cal F}_{\mu}/B_{\|}^{*},\frac{}{} 
{\cal H} \right\}_{\rm gc} \nonumber \\
 &  &+\; B_{\|}^{*}\;\left\{ \delta{\cal S}\,{\cal H},\frac{}{} {\cal F}_{\mu}/B_{\|}^{*} \right\}_{\rm gc} \nonumber \\
  &  &+\; \pd{}{Z^{\alpha}}\left({\cal F}_{\mu}{\cal H}\;\frac{e}{c}\,\delta{\bf A}^{*}\bdot \left\{\vb{\sf X},\frac{}{} Z^{\alpha}\right\}_{\rm gc} \right)
  \label{eq:delta_FH}
 \end{eqnarray}
 and the last term in Eq.~\eqref{eq:delta_FH} vanishes when integrated upon $w$ through the identity \eqref{eq:FH_id}.
 
  Using the variations $(\delta\Phi^{*}, \delta{\bf A}^{*})$ given by Eqs.~\eqref{eq:delta_Phi} and \eqref{eq:delta_A},  we find
 \begin{eqnarray} 
 e\,\delta\Phi^{*} - \frac{e}{c}\,\delta{\bf A}^{*}\bdot\frac{d_{\rm gc}{\bf X}}{dt} & = & e\,\delta\Phi - \frac{e}{c}\,\delta{\bf A}\bdot\frac{d_{\rm gc}{\bf X}}{dt} 
 \label{eq:PBgc_B} \\
  &  &-\; \delta{\bf B}\bdot\left( \vb{\mu}_{\rm gc} \;+\; \vb{\pi}_{\rm gc}\btimes\frac{{\sf p}_{\|}\,\bhat}{mc} \right), \nonumber
  \end{eqnarray}
  where we used the definitions found in Eq~\eqref{eq:Lgc_B}. Hence, Eq.~\eqref{eq:delta_Lgc_def} becomes
  \begin{eqnarray}
  \delta{\cal L}_{\rm gc} & = & \frac{1}{4\pi} \left( \delta{\bf E}\bdot{\bf E} -\frac{}{} \delta{\bf B}\bdot{\bf H} \right) \nonumber \\
    &  &- \int  \left( e\,\delta\Phi - \frac{e}{c}\,\delta{\bf A}\bdot\frac{d_{\rm gc}{\bf X}}{dt}\right) {\cal F}_{\mu} \,d^{3}P \nonumber \\
   &  &-\; \int  B_{\|}^{*}\;\delta{\cal S}\, \left\{{\cal F}_{\mu}/B_{\|}^{*},\frac{}{} {\cal H} \right\}_{\rm gc} \,d^{3}P \nonumber \\
    &  &-\; \int  B_{\|}^{*}\;\left\{ \delta{\cal S}\,{\cal H},\frac{}{} {\cal F}_{\mu}/B_{\|}^{*} \right\}_{\rm gc} \,d^{3}P, 
 \end{eqnarray}
 where ${\bf H}$ is defined by Eq.~\eqref{eq:Hgc_field}.

Next, using the phase-space divergence form of the guiding-center Poisson bracket \eqref{eq:gcPB_ext}, we find
\begin{eqnarray*} 
B_{\|}^{*}\left\{ \delta{\cal S}\,{\cal H},\frac{}{} {\cal F}_{\mu}/B_{\|}^{*} \right\}_{\rm gc} & = & -\;\pd{}{Z^{\alpha}}\left({\cal F}_{\mu}\;\left\{ Z^{\alpha},\frac{}{} \delta{\cal S}\;
{\cal H} \right\}_{\rm gc} \right) \\
 & = & -\;\pd{}{Z^{\alpha}}\left({\cal F}_{\mu}\,{\cal H}\;\left\{ Z^{\alpha},\frac{}{} \delta{\cal S}\right\}_{\rm gc} \right) \\
  &  &-\;\pd{}{Z^{\alpha}}\left({\cal F}_{\mu}\,\delta{\cal S}\;\frac{d_{\rm gc}Z^{\alpha}}{dt}\right),
\end{eqnarray*}
which yields
\begin{eqnarray*}
 &  &-\; \int  B_{\|}^{*}\;\left\{ \delta{\cal S}\,{\cal H},\frac{}{} {\cal F}_{\mu}/B_{\|}^{*} \right\}_{\rm gc} \,d^{3}P \\
  & &\hspace*{0.5in}=\; \pd{}{t} \left(\int  {\cal F}_{\mu}\;\delta{\cal S}\,d^{3}P \right) \\
 &  &\hspace*{0.7in}+\; \nabla\bdot\left(\int  {\cal F}_{\mu}\;\delta{\cal S}\;\frac{d_{\rm gc}{\bf X}}{dt} \,d^{3}P \right),
 \end{eqnarray*}
 where we used the identity \eqref{eq:FH_id}. 
 
Lastly, using Eq.~\eqref{eq:delta_EB}, we obtain the final form for the Eulerian variation of the guiding-center Lagrangian density:
\begin{eqnarray}
\delta{\cal L}_{\rm gc} & \equiv & -\,\int  B_{\|}^{*}\;\delta{\cal S}\, \{{\cal F}_{\mu}/B_{\|}^{*},\; {\cal H}\}_{\rm gc} \,d^{3}P \nonumber \\
 &  &+\; \frac{\delta\Phi}{4\pi} \left( \nabla\bdot{\bf E} \;-\frac{}{} 4\pi\,\varrho_{\rm gc} \right) \nonumber \\
  &  &+\; \frac{\delta{\bf A}}{4\pi}\bdot\left( \frac{1}{c}\pd{\bf E}{t} - \nabla\btimes{\bf H}  + \frac{4\pi}{c}\,{\bf J}_{\rm gc} \right) \nonumber \\
   &  &+\; \pd{\delta{\cal J}}{t} \;+\; \nabla\bdot\delta\vb{\Gamma},
   \label{eq:delta_Lgc}
\end{eqnarray}
where the space-time derivatives are expressed in terms of the Noether components
\begin{eqnarray}
\delta{\cal J} & \equiv & \int  \delta{\cal S}\;{\cal F}_{\mu}\;d^{3}P \;-\; \frac{{\bf E}\bdot\delta{\bf A}}{4\pi\,c}, \label{eq:delta_Lambda} \\
\delta\vb{\Gamma} & \equiv &  \int  \delta{\cal S}\;{\cal F}_{\mu}\;\frac{d_{\rm gc}{\bf X}}{dt}\;d^{3}P \nonumber \\
 &  &-\; \frac{1}{4\pi} \left( \delta\Phi\,{\bf E} \;+\frac{}{} \delta{\bf A}\btimes{\bf H} \right). \label{eq:delta_Gamma}
\end{eqnarray}
When the Eulerian variation \eqref{eq:delta_Lgc_def} is inserted into the Euler action variation 
\[ \delta{\cal A}_{\rm gc}^{\rm E} \;\equiv\; \int  \;\delta{\cal L}_{\rm gc}\;d^{3}x\;dt \;\equiv\; 0, \]
we obtain
\begin{eqnarray}
\delta{\cal A}_{\rm gc}^{\rm E} & = & -\,\int  B_{\|}^{*}\,\delta{\cal S}\;\{ {\cal F}_{\mu}/B_{\|}^{*},\; {\cal H}\}_{\rm gc}\,d^{6}Z\,d\mu
 \label{delta_Agc_E} \\
 &  &+\; \int \frac{\delta\Phi}{4\pi} \left( \nabla\bdot{\bf E} \;-\frac{}{} 4\pi\,\varrho_{\rm gc} \right) d^{3}x\;dt\nonumber \\
  &  &+\; \int \frac{\delta{\bf A}}{4\pi}\bdot\left( \frac{1}{c}\pd{\bf E}{t} - \nabla\btimes{\bf H}  + \frac{4\pi}{c}\,{\bf J}_{\rm gc} \right) d^{3}x\;dt,
\nonumber
 \end{eqnarray}
 where the Noether terms in Eq.~\eqref{eq:delta_Lgc} vanish upon space-time integration. Stationarity with respect to the variations $(\delta\Phi,\delta{\bf A})$ yields the guiding-center Maxwell equations \eqref{eq:div_E}-\eqref{eq:curl_B}, while stationarity with respect to the variation $\delta{\cal S}$ yields the extended guiding-center Vlasov equation 
 $B_{\|}^{*}\{{\cal F}_{\mu}/B_{\|}^{*},\,{\cal H}\}_{\rm gc} = 0$, which becomes the phase-space divergence form \eqref{eq:ext_PB_HF} of the guiding-center Vlasov equation when integrated over $w$.
 
 \section{\label{sec:Noether}Guiding-center Noether Equation}
 
The Noether method plays an important role in deriving exact conservation laws from symmetries of the Lagrangian density \cite{Brizard_2000,Brizard_2005,Brizard_Lag}.  Once the guiding-center Vlasov-Maxwell equations are inserted into the Eulerian variation \eqref{eq:delta_Lgc} of the guiding-center Lagrangian density, we obtain the guiding-center Noether equation
\begin{equation}
\delta{\cal L}_{\rm gc} \;=\; \pd{\delta{\cal J}}{t} \;+\; \nabla\bdot\delta\vb{\Gamma},
\label{eq:Noether_gc}
\end{equation}
where the Noether components are defined in Eqs.~\eqref{eq:delta_Lambda}-\eqref{eq:delta_Gamma}. 

The energy-momentum conservation laws are associated with Lagrangian symmetry under space-time translations, which are generated by the scalar field
\begin{equation}
\delta{\cal S} \;\equiv\; {\bf P}\bdot\delta{\bf x} \;-\; w\,\delta t,
\label{eq:deltaS_xt}
\end{equation}
where ${\bf P} \equiv (e/c)\,{\bf A}^{*}$ represents the canonical guiding-center momentum appearing as the spatial component of the guiding-center one-form \eqref{eq:Gamma_gc}, and $-\,w$ appears as its time component. The variations $(\delta\Phi,\delta{\bf A})$, on the other hand, are expressed as
\begin{equation}
\left. \begin{array}{rcl}
\delta\Phi & \equiv & \delta{\bf x}\bdot{\bf E} \;+\; c^{-1}\partial\delta\chi/\partial t \\
 &  & \\
 \delta{\bf A} & \equiv & c\,\delta t\;{\bf E} \;+\; \delta{\bf x}\btimes{\bf B} \;-\; \nabla\delta\chi
 \end{array} \right\},
 \label{eq:delta_PhiA_xt}
 \end{equation}
 where the virtual gauge field $\delta\chi$ is defined as
 \begin{equation}
 \delta\chi \;\equiv\; {\bf A}\bdot\delta{\bf x} \;-\; \Phi\;c\,\delta t.
 \label{eq:delta_chi}
 \end{equation}
 Lastly, the guiding-center energy-momentum conservation laws are derived from the Noether equation \eqref{eq:Noether_gc} by considering the variation
 \begin{equation}
 \delta{\cal L}_{\rm gc} \;\equiv\; -\,\left(\delta t\;\pd{}{t} + \delta{\bf x}\bdot\nabla\right) {\cal L}_{\rm M},
 \label{eq:delta_Lgc_xt}
 \end{equation}
 where ${\cal L}_{\rm M} \equiv (|{\bf E}|^{2} - |{\bf B}|^{2})/8\pi$ is the Maxwell Lagrangian density and the guiding-center Vlasov term does not appear in
 Eq.~\eqref{eq:delta_Lgc_xt} because of the identity \eqref{eq:FH_id}.
 
 \subsection{Guiding-center energy conservation law}
 
 By considering an infinitesimal time translation $\delta t$, the Noether equation \eqref{eq:Noether_gc} yields the {\it primitive} energy conservation law
 \begin{eqnarray}
 0 & = & \pd{}{t} \left[ \left(\int  w\,{\cal F}_{\mu}\;d^{3}P \right) + \frac{1}{8\pi} \left( |{\bf E}|^{2} +\frac{}{} |{\bf B}|^{2}\right) \right] 
 \label{eq:energy_gc_prime} \\
  &  &+ \nabla\bdot\left[ \left(\int  w\,{\cal F}_{\mu}\,\frac{d_{\rm gc}{\bf X}}{dt}\;d^{3}P \right) \;+\; \frac{c}{4\pi}\;{\bf E}\btimes{\bf H} \right] \nonumber \\
   &  &+ \pd{}{t} \left( \frac{\bf E}{4\pi}\bdot \nabla\Phi \right) + \nabla\bdot\left( \nabla\Phi\btimes\frac{c\,{\bf H}}{4\pi} - \frac{\bf E}{4\pi}\;\pd{\Phi}{t}
   \right), \nonumber
 \end{eqnarray}
 where the $\Phi$-terms are associated with the gauge term \eqref{eq:delta_chi}: $\Phi \equiv -\,c^{-1}\delta\chi/\delta t$. These gauge terms are actually important in deriving a gauge-independent guiding-center energy conservation law.
 
 First, we use the constraint \eqref{eq:Fgc_ext} on the extended guiding-center Vlasov distribution to obtain
 \[ \int \left(w, w\frac{d_{\rm gc}{\bf X}}{dt}\right)  {\cal F}_{\mu}\,d^{3}P  \equiv \int  \left(H_{\rm gc}, H_{\rm gc}\frac{d_{\rm gc}{\bf X}}{dt}\right) F_{\mu}\,d{\sf p}_{\|}d\mu, 
\]
 where the guiding-center Hamiltonian is
 \[ H_{\rm gc} \;=\; \left(\mu B \;+\; \frac{p_{\|}^{2}}{2m}\right) + e\,\Phi \;\equiv\; K_{\rm gc} + e\,\Phi. \]
 Next, we write
 \begin{equation}
 \pd{}{t} \left( \frac{\bf E}{4\pi}\bdot \nabla\Phi \right) = \pd{}{t}\left[\nabla\bdot\left(\Phi\,\frac{\bf E}{4\pi}\right) - \frac{\Phi}{4\pi}\left(\nabla\bdot{\bf E}\right) \right],
 \label{eq:primitive_one}
 \end{equation}
 and
 \begin{eqnarray}
 &  & \nabla\bdot\left( \nabla\Phi\btimes\frac{c\,{\bf H}}{4\pi} \;-\; \frac{\bf E}{4\pi}\;\pd{\Phi}{t}\right) \label{eq:primitive_two}  \\
  & = & -\;\nabla\bdot\left[\pd{}{t}\left(\Phi\;\frac{\bf E}{4\pi}\right) \;-\; \frac{\Phi}{4\pi}\left( \pd{\bf E}{t} \;-\; c\,\nabla\btimes{\bf H} \right) \right],
\nonumber
\end{eqnarray}
so that, when combining Eqs.~\eqref{eq:primitive_one}-\eqref{eq:primitive_two}, and using the guiding-center Maxwell equations \eqref{eq:div_E}-\eqref{eq:curl_B}, we obtain
\begin{eqnarray*}
 &  & \pd{}{t} \left( \frac{\bf E}{4\pi}\bdot \nabla\Phi \right) \;+\; \nabla\bdot\left( \nabla\Phi\btimes\frac{c\,{\bf H}}{4\pi} \;-\; \frac{\bf E}{4\pi}\;\pd{\Phi}{t}
   \right) \nonumber \\
  & = & -\pd{}{t} \left[ \frac{\Phi}{4\pi}\left(\nabla\bdot{\bf E}\right) \right] + \nabla\bdot\left[\frac{\Phi}{4\pi}\left( \pd{\bf E}{t} - c\,\nabla\btimes{\bf H} \right) \right] \\
   & \equiv & -\int \;e  \left[\pd{}{t}\left( \Phi\frac{}{}F_{\mu} \right) + \nabla\bdot\left(\frac{d_{\rm gc}{\bf X}}{dt}\; \Phi\,F_{\mu}\right)\right]\,d{\sf p}_{\|}d\mu.
  \end{eqnarray*}
  Hence, by combining Eqs.~\eqref{eq:primitive_one}-\eqref{eq:primitive_two} into Eq.~\eqref{eq:energy_gc_prime}, we obtain the guiding-center energy conservation law
  \begin{equation}
  \pd{{\cal E}_{\rm gc}}{t} \;+\; \nabla\bdot{\bf S}_{\rm gc} \;=\; 0,
  \label{eq:energy_gc}
  \end{equation}
 where the guiding-center energy density ${\cal E}_{\rm gc}$ and the guiding-center energy-density flux ${\bf S}_{\rm gc}$ are defined as
  \begin{eqnarray}
{\cal E}_{\rm gc} & \equiv & \int F_{\mu}\;K_{\rm gc}\,d{\sf p}_{\|}d\mu \;+\; \frac{1}{8\pi} \left( |{\bf E}|^{2} \;+\frac{}{} |{\bf B}|^{2}\right),
 \label{eq:E_gc} \\
{\bf S}_{\rm gc} & \equiv & \int F_{\mu}\;K_{\rm gc}\,\frac{d_{\rm gc}{\bf X}}{dt}\,d{\sf p}_{\|}d\mu \;+\; \frac{c}{4\pi}\;{\bf E}\btimes{\bf H}.
\label{eq:S_gc}
 \end{eqnarray}
We note here that guiding-center magnetization (including its polarization contribution) does not appear in the guiding-center energy density \eqref{eq:E_gc} but only in the guiding-center energy-density flux \eqref{eq:S_gc}. 
 
To prove the guiding-center energy conservation law \eqref{eq:energy_gc}, we begin with
\begin{eqnarray}
\pd{{\cal E}_{\rm gc}}{t} & = & \int  \left[ K_{\rm gc}\;\pd{F_{\mu}}{t} \;+\; F_{\mu}\;\left(\mu\;\pd{B}{t} \right) \right]\,d{\sf p}_{\|}d\mu \nonumber \\
 &  &+\; \frac{1}{4\pi}\left( {\bf E}\bdot\pd{\bf E}{t} \;+\; {\bf B}\bdot\pd{\bf B}{t} \right) 
 \label{eq:Egc_1} \\
  & = & -\;\nabla\bdot\left( \int K_{\rm gc}\,\frac{d_{\rm gc}{\bf X}}{dt}\; F_{\mu}\,d{\sf p}_{\|}d\mu \right) \nonumber \\
   &  &+\; \int \left( \frac{d_{\rm gc}{\bf X}}{dt}\bdot \mu\,\nabla B \;+\; \frac{{\sf p}_{\|}}{m}\;\frac{d_{\rm gc}p_{\|}}{dt} \right) F_{\mu}\,d{\sf p}_{\|}d\mu \nonumber \\
    &  &+\; \frac{\bf E}{4\pi}\bdot\left(c\,\nabla\btimes{\bf H} \;-\; 4\pi\,\int  e\;\frac{d_{\rm gc}{\bf X}}{dt}\;F_{\mu}\,d{\sf p}_{\|}d\mu \right) \nonumber \\
     &  &-\; \left( \frac{\bf B}{4\pi} \;+\; \int \mu\,\bhat\;F_{\mu}\,d{\sf p}_{\|}d\mu \right)\bdot c\,\nabla\btimes{\bf E}, \nonumber 
 \end{eqnarray}
 where we made use of the phase-space divergence form \eqref{eq:gcV_div} of the guiding-center Vlasov equation and the guiding-center Maxwell equation \eqref{eq:curl_B}. Next, we use Eqs.~\eqref{eq:Xgc_dot}-\eqref{eq:pgc_dot} to obtain the relation
\begin{eqnarray*} 
\frac{{\sf p}_{\|}}{m}\;\frac{d_{\rm gc}p_{\|}}{dt} & = & e\,{\bf E}^{*}\bdot\left(\frac{{\sf p}_{\|}}{m}\;\frac{{\bf B}^{*}}{B_{\|}^{*}} \right) \;=\; e\,{\bf E}^{*}\bdot
\frac{d_{\rm gc}{\bf X}}{dt} \\
 & = & \left( e\,{\bf E} \;-\; \mu\,\nabla B \;-\; {\sf p}_{\|}\,\pd{\bhat}{t}\right)\bdot\frac{d_{\rm gc}{\bf X}}{dt},
 \end{eqnarray*}
 so that several terms in Eq.~\eqref{eq:Egc_1} cancel out. With these cancellations, and using the definition \eqref{eq:Hgc_field} as well as the Maxwell equation \eqref{eq:curl_E},  we recover Eq.~\eqref{eq:energy_gc} from Eq.~\eqref{eq:Egc_1}.
 
 \subsection{Guiding-center momentum conservation law}
 
 By considering an infinitesimal spatial translation $\delta{\bf x}$, the Noether equation \eqref{eq:Noether_gc} yields the {\it primitive} momentum conservation law
 \begin{eqnarray}
 0 & = & \pd{}{t} \left( \int  {\bf P}\,F_{\mu}\,d{\sf p}_{\|}d\mu \;+\; \frac{{\bf E}\btimes{\bf B}}{4\pi\,c}\right) 
 \label{eq:momentum_gc_prime} \\
  &  &+\; \nabla\bdot\left[ \int  \frac{d_{\rm gc}{\bf X}}{dt}\,{\bf P}\,F_{\mu}\,d{\sf p}_{\|}d\mu \;-\; \left( \frac{{\bf E}{\bf E}}{4\pi} -
  \frac{|{\bf E}|^{2}}{8\pi}\;{\bf I} \right) \right. \nonumber \\
   &  &\left.-\; \left( \frac{{\bf B}{\bf B}}{4\pi} - \frac{|{\bf B}|^{2}}{8\pi}\;{\bf I} \right) \;+\; \left( {\bf B}\,{\bf M}_{\rm gc} \;-\frac{}{} {\bf B}\bdot{\bf M}_{\rm gc}\,{\bf I}
   \right) \right] \nonumber \\
    &  &+ \pd{}{t} \left( \frac{{\bf E}\bdot\nabla{\bf A}}{4\pi\,c} \right) - \nabla\bdot\left( \frac{\bf E}{4\pi\,c}\;\pd{\bf A}{t} +
    \nabla\btimes{\bf H}\;\frac{\bf A}{4\pi} \right), \nonumber
 \end{eqnarray}
where the ${\bf A}$-terms are associated with the gauge term \eqref{eq:delta_chi}: ${\bf A} \equiv \delta\chi/\delta{\bf x}$. These gauge terms are again 
important in deriving a gauge-independent guiding-center momentum conservation law.

First, we use the guiding-center Maxwell equations \eqref{eq:div_E}-\eqref{eq:curl_B} to obtain
\begin{eqnarray}
\pd{}{t} \left( \frac{{\bf E}\bdot\nabla{\bf A}}{4\pi\,c} \right) & = & \pd{}{t}\left[\nabla\bdot\left( \frac{{\bf E}{\bf A}}{4\pi\,c}\right) \right. \nonumber \\
 &  &\left.\hspace*{0.3in}-\;\frac{e}{c}\,{\bf A}\; \left(\int  F_{\mu}\,d{\sf p}_{\|}d\mu \right)\right], \label{eq:momentum_one} 
 \end{eqnarray}
 and
 \begin{eqnarray}
 &  &-\;\nabla\bdot\left( \frac{\bf E}{4\pi\,c}\;\pd{\bf A}{t} + \nabla\btimes{\bf H}\;\frac{\bf A}{4\pi} \right) \label{eq:momentum_two} \\
  & = & -\;\nabla\bdot\left[ \pd{}{t}\left( \frac{{\bf E}{\bf A}}{4\pi\,c}\right) \;+\; \frac{e}{c}\,{\bf A}\; \left(\int \frac{d_{\rm gc}{\bf X}}{dt}\,F_{\mu}\,d{\sf p}_{\|}d\mu\right)\right],
  \nonumber
  \end{eqnarray}
  which yield gauge cancellations in Eq.~\eqref{eq:momentum_gc_prime}, so that Eq.~\eqref{eq:momentum_gc_prime} becomes
  \begin{eqnarray}
  0 & = & \pd{}{t} \left(\int {\sf p}_{\|}\,\bhat\;F_{\mu}\,d{\sf p}_{\|}d\mu \;+\; \frac{{\bf E}\btimes{\bf B}}{4\pi\,c} \right) 
     \label{eq:momentum_gc} \\
   &  &+\; \nabla\bdot\left[ \int \frac{d_{\rm gc}{\bf X}}{dt}\,{\sf p}_{\|}\,\bhat\,F_{\mu}\,d{\sf p}_{\|}d\mu \;-\; \left( \frac{{\bf E}{\bf E}}{4\pi} -
  \frac{|{\bf E}|^{2}}{8\pi}\;{\bf I} \right) \right. \nonumber \\
   &  &\left.-\; \left( \frac{{\bf B}{\bf B}}{4\pi} - \frac{|{\bf B}|^{2}}{8\pi}\;{\bf I} \right) \;+\; \left( {\bf B}\,{\bf M}_{\rm gc} \;-\frac{}{} {\bf B}\bdot{\bf M}_{\rm gc}\,{\bf I}
   \right) \right]. \nonumber
  \end{eqnarray}
  
  Next, by using the definition \eqref{eq:Mgc} for the guiding-center magnetization, we obtain the relation
 \begin{eqnarray}
 &  & {\bf B}\,{\bf M}_{\rm gc} \;-\frac{}{} {\bf B}\bdot{\bf M}_{\rm gc}\,{\bf I} 
 \label{eq:BM_id} \\
  & = &  \int  \left[ {\bf B}\left( \vb{\mu}_{\rm gc} + \vb{\pi}_{\rm gc}\btimes\frac{{\sf p}_{\|}\bhat}{mc}\right) - \left({\bf B}\bdot\vb{\mu}_{\rm gc}\right)\,{\bf I} 
  \right]\;F_{\mu}\,d{\sf p}_{\|}d\mu \nonumber \\
   & \equiv & \int  \left[ \mu\,B \left({\bf I} - \bhat\bhat\right) \;+\; {\sf p}_{\|}\;\bhat \left( \frac{d_{\rm gc}{\bf X}}{dt}\right)_{\bot} \right]\,F_{\mu}\,d{\sf p}_{\|}d\mu,
   \nonumber
 \end{eqnarray}
 where the perpendicular guiding-center drift velocity is defined as
\begin{equation}
\left(\frac{d_{\rm gc}{\bf X}}{dt}\right)_{\bot} \;\equiv\; \left(\bhat\btimes\frac{d_{\rm gc}{\bf X}}{dt}\right)\btimes\bhat \;=\; \frac{d_{\rm gc}{\bf X}}{dt} \;-\; 
\frac{{\sf p}_{\|}}{m}\,\bhat.
\end{equation}

Hence, Eq.~\eqref{eq:momentum_gc} becomes the guiding-center momentum conservation law
 \begin{equation}
 \pd{{\bf P}_{\rm gc}}{t} \;+\; \nabla\bdot{\sf T}_{\rm gc} \;=\; 0,
 \label{eq:momentum_gc_final}
 \end{equation}
 where the guiding-center momentum density ${\bf P}_{\rm gc}$ is defined as
 \begin{equation}
 {\bf P}_{\rm gc} \;\equiv\; \int {\sf p}_{\|}\,\bhat\;F_{\mu}\,d{\sf p}_{\|}d\mu \;+\; \frac{{\bf E}\btimes{\bf B}}{4\pi\,c}, \label{eq:P_gc}
 \end{equation}
 and the guiding-center stress tensor 
 \begin{equation}
 {\sf T}_{\rm gc} \;\equiv\; {\sf T}_{\rm M} \;+\; {\sf T}_{\rm gcV} 
 \label{eq:Tgc_def}
 \end{equation}
 is defined as the sum of the symmetric Maxwell tensor 
 \begin{equation}
 {\sf T}_{\rm M} \equiv \left( |{\bf E}|^{2} \;+\frac{}{} |{\bf B}|^{2}\right) \frac{\bf I}{8\pi} - \frac{1}{4\pi} \left({\bf E}{\bf E} \;+\frac{}{} {\bf B}{\bf B}
\right), \label{eq:T_Maxwell}
\end{equation}
and the symmetric guiding-center Vlasov stress tensor
\begin{eqnarray}
 {\sf T}_{\rm gcV} & \equiv & \int \left[ {\sf p}_{\|} \left( \frac{d_{\rm gc}{\bf X}}{dt} \right)_{\bot}\,\bhat \;+\; {\sf p}_{\|}\,\bhat\,
 \left( \frac{d_{\rm gc}{\bf X}}{dt}\right)_{\bot} \right. \nonumber \\
  &  &\left.+\; \frac{{\sf p}_{\|}^{2}}{m}\; \bhat\,\bhat  \;+\; \mu\,B\;\left({\bf I} - \bhat\bhat\right) \right]F_{\mu}\,d{\sf p}_{\|}d\mu,
  \label{eq:Tgc_V}
\end{eqnarray}
with nonvanishing off-diagonal components, where we wrote
\[  {\sf p}_{\|}\;\frac{d_{\rm gc}{\bf X}}{dt}\,\bhat \;\equiv\; {\sf p}_{\|} \left( \frac{d_{\rm gc}{\bf X}}{dt} \right)_{\bot}\,\bhat \;+\; \frac{{\sf p}_{\|}^{2}}{m}\;\bhat\bhat. \]
Here, we note that the symmetry of the guiding-center Vlasov stress tensor \eqref{eq:Tgc_V} is an explicit consequence of the polarization contribution to the guiding-center magnetization used to derive the relation \eqref{eq:BM_id}. In addition, we note that the guiding-center Vlasov stress tensor \eqref{eq:Tgc_V} includes the CGL pressure tensor
\begin{equation}
{\sf P}_{\rm CGL} \equiv  \int \left[ \frac{{\sf p}_{\|}^{2}}{m}\;\bhat\,\bhat +\frac{}{} \mu\,B\;\left({\bf I} - \bhat\bhat\right) \right]\;F_{\mu}\,d{\sf p}_{\|}d\mu.
\label{eq:P_CGL}
\end{equation}
We also note that the explicit symmetric form of guiding-center stress tensor \eqref{eq:Tgc_def} is in contrast to the standard expressions found in previous works \cite{Similon_1985}, where the symmetry is only implicitly assumed.

To prove the guiding-center momentum conservation law \eqref{eq:momentum_gc_final}, we begin with
\begin{widetext}
\begin{eqnarray}
\pd{{\bf P}_{\rm gc}}{t} & = & \int \left[ {\sf p}_{\|}\,\bhat\;\pd{F_{\mu}}{t} \;+\; F_{\mu}\;{\sf p}_{\|}\,\pd{\bhat}{t} \right]\,d{\sf p}_{\|}d\mu \;+\; \frac{1}{4\pi\,c} \left( \pd{\bf E}{t}\btimes{\bf B} \;+\; {\bf E}\btimes\pd{\bf B}{t} \right) \label{eq:momentum_one} \\
  & = & -\;\nabla\bdot{\sf T}_{\rm gc} \;-\; \nabla{\bf B}\bdot{\bf M}_{\rm gc} + \int \left[ \frac{d_{\rm gc}(p_{\|}\bhat)}{dt} - e\,\left( {\bf E} + \frac{1}{c}\,\frac{d_{\rm gc}{\bf X}}{dt}\btimes{\bf B}\right) \right] F_{\mu}\,d{\sf p}_{\|}d\mu, \nonumber
 \end{eqnarray}
\end{widetext} 
where the guiding-center Vlasov-Maxwell equations were used to obtain the second expression. Next, we use the Euler-Lagrange equation
 \eqref{eq:EL_X} to obtain
 \begin{eqnarray*}
 \frac{d_{\rm gc}(p_{\|}\bhat)}{dt} & = & e\,\left( {\bf E} + \frac{1}{c}\,\frac{d_{\rm gc}{\bf X}}{dt}\btimes{\bf B}\right) \\
  &  &-\; \mu\,\nabla B \;+\; {\sf p}_{\|}\,\nabla\bhat\bdot\frac{d_{\rm gc}{\bf X}}{dt},
 \end{eqnarray*}
 and we use the definition \eqref{eq:Mgc} of the guiding-center magnetization ${\bf M}_{\rm gc}$ to obtain
 \begin{eqnarray*}
-\; \nabla{\bf B}\bdot{\bf M}_{\rm gc} & = & \int \nabla{\bf B}\bdot\left[ \mu\,\bhat - \frac{{\sf p}_{\|}}{B}\left(\frac{d_{\rm gc}{\bf X}}{dt}
\right)_{\bot} \right] F_{\mu}\,d{\sf p}_{\|}d\mu \\
 & = & \int \left( \mu\,\nabla B - {\sf p}_{\|}\,\nabla\bhat\bdot\frac{d_{\rm gc}{\bf X}}{dt} \right) F_{\mu}\,d{\sf p}_{\|}d\mu
\end{eqnarray*}
so that we recover the guiding-center momentum conservation law \eqref{eq:momentum_gc_final}.

\subsection{Guiding-center toroidal angular momentum conservation law}

The conservation law for toroidal angular momentum is particularly relevant to plasmas confined by axisymmetric magnetic fields, which are generically expressed in terms of the mixed representation
\begin{equation}
{\bf B} \;\equiv\; B_{\varphi}(\psi)\;\nabla\varphi \;+\; \nabla\varphi\btimes\nabla\psi,
\label{eq:B_axis}
\end{equation}
where $\varphi$ denotes the toroidal angle, which is an ignorable angle in axisymmetric magnetized plasmas, $\psi$ denotes the poloidal magnetic flux, and the toroidal magnetic-field (covariant) component $B_{\varphi}(\psi)$ is a flux function. Note that the magnetic field 
\eqref{eq:B_axis} satisfies the relations
\begin{equation}
\left. \begin{array}{rcl}
{\bf B}\bdot\nabla\psi & = & 0 \\
 &  & \\
 {\bf B}\btimes\partial{\bf x}/\partial\varphi & = & \nabla\psi
 \end{array} \right\}.
 \label{eq:B_psi}
 \end{equation}
 In this Section, we will use the magnetic coordinates $(\psi,\vartheta,\varphi)$ to represent the guiding-center position ${\bf X}$, where $\vartheta$ denotes the poloidal angle, with magnetic-coordinate Jacobian $(\nabla\psi\btimes\nabla\vartheta\bdot\nabla\varphi)^{-1} \equiv 1/B^{\vartheta}$ expressed in terms of the poloidal contravariant component $B^{\vartheta} \equiv {\bf B}\bdot\nabla\vartheta$.
 
 We can now derive the toroidal angular momentum conservation law for the guiding-center Vlasov-Maxwell equations either directly from the guiding-center momentum conservation law \eqref{eq:momentum_gc_final} or from the guiding-center Noether equation \eqref{eq:Noether_gc} with $\delta{\bf x} \equiv 
 \delta\varphi\;\partial{\bf x}/\partial\varphi$. Here, we begin with Eq.~\eqref{eq:momentum_gc_final} and, taking the dot product with $\partial{\bf x}/\partial\varphi$, we obtain the primitive guiding-center toroidal angular momentum conservation law
 \begin{equation}
 \pd{P_{{\rm gc}\varphi}}{t} \;+\; \nabla\bdot\left({\sf T}_{\rm gc}\bdot\pd{\bf x}{\varphi}\right) \;=\; \nabla\left(\pd{\bf x}{\varphi}\right):{\sf T}_{\rm gc}^{\top} \;\equiv\; 0,
 \label{eq:toroidal_prime}
 \end{equation}
 where $P_{{\rm gc}\varphi} \equiv {\bf P}_{\rm gc}\bdot\partial{\bf x}/\partial\varphi$ denotes toroidal covariant component of the guiding-center momentum density \eqref{eq:P_gc} and the right side vanishes since the guiding-center stress tensor \eqref{eq:Tgc_def} is symmetric, i.e., ${\sf T}_{\rm gc}^{\top} \equiv 
 {\sf T}_{\rm gc}$, while the dyadic tensor $\nabla(\partial{\bf x}/\partial\varphi)$ is antisymmetric.
 
 More explicitly, we begin with the primitive guiding-center toroidal angular momentum conservation law \eqref{eq:toroidal_prime}, which can be written as
 \begin{eqnarray}
 0 & = & \pd{}{t}\left( \int  F_{\mu}\,{\sf p}_{\|}\,b_{\varphi}\,d{\sf p}_{\|}d\mu \;+\; \frac{E^{\psi}}{4\pi\,c} \right) \label{eq:tormag_prim} \\
  &  &+\; \nabla\bdot\left[ \int \frac{d_{\rm gc}{\bf X}}{dt}\;F_{\mu}\,{\sf p}_{\|}\,b_{\varphi}\,d{\sf p}_{\|}d\mu \;-\; \frac{\bf E}{4\pi\,c}\;\pd{\psi}{t}  \right. \nonumber \\
   &  &\left.- \frac{\bf B}{4\pi} {\bf H}\bdot\pd{\bf x}{\varphi} + \pd{\bf x}{\varphi} \left( \frac{1}{8\pi}\left(|{\bf E}|^{2} + |{\bf B}|^{2}\right) - {\bf B}\bdot{\bf M}_{\rm gc}
   \right) \right],
   \nonumber
   \end{eqnarray}
where we used ${\bf E}\btimes{\bf B}\bdot\partial{\bf x}/\partial\varphi = {\bf E}\bdot\nabla\psi \equiv E^{\psi}$ and $E_{\varphi} \equiv c^{-1}\partial\psi/\partial t$ (assuming axisymmetry: $\partial\Phi/\partial\varphi \equiv 0$ and using ${\bf A}\bdot\partial{\bf x}/\partial\varphi \equiv -\,\psi$). Next, we use the identity
\begin{eqnarray*}
-\; \frac{\bf E}{4\pi\,c}\;\pd{\psi}{t} & = & -\;\pd{}{t}\left(\frac{\psi}{4\pi c}\;{\bf E}\right) \;+\; \frac{\psi}{4\pi}\;\nabla\btimes{\bf H} \\
 &  &-\;\frac{e}{c}\,\psi\;\left(\int  \frac{d_{\rm gc}{\bf X}}{dt}\;F_{\mu}\,d{\sf p}_{\|}d\mu\right),
\end{eqnarray*}
where we used the guiding-center Maxwell equation \eqref{eq:curl_B}, so that the divergence of the first term on the right side yields
\[ \nabla\bdot\left( -\; \frac{\bf E}{4\pi\,c}\;\pd{\psi}{t} \right) \;=\; -\;\pd{}{t}\left(\frac{\psi}{4\pi c}\;\nabla\bdot{\bf E} \;+\; \frac{E^{\psi}}{4\pi\,c} \right), \]
while the divergence of the second term yields
\begin{eqnarray*} 
\nabla\bdot\left(\frac{\psi}{4\pi}\nabla\btimes{\bf H}\right) & = & \nabla\bdot\left(\frac{\bf H}{4\pi}\btimes\nabla\psi\right) \\
 & = & \nabla\bdot\left[ \frac{\bf B}{4\pi}{\bf H}\bdot\pd{\bf x}{\varphi} - \pd{\bf x}{\varphi} \left(\frac{{\bf B}\bdot{\bf H}}{4\pi}\right) \right].
 \end{eqnarray*}
 These divergence terms lead to cancellations in Eq.~\eqref{eq:tormag_prim} so that, by combining the remaining terms, we obtain the guiding-center toroidal angular momentum conservation law
 \begin{eqnarray}
  & &\pd{}{t}\left(\int  F_{\mu}\,P_{\varphi}\,d{\sf p}_{\|}d\mu \right) + \nabla\bdot\left( \int  \frac{d_{\rm gc}{\bf X}}{dt}F_{\mu}\,P_{\varphi}\,d{\sf p}_{\|}d\mu \right) \nonumber \\
   & &\hspace*{0.5in}=\; -\,\pd{{\cal L}_{\rm M}}{\varphi}  \;\equiv\; 0,
 \label{eq:P_phi_dot}
\end{eqnarray}
where $P_{\varphi} \equiv (e/c)\,{\bf A}^{*}\bdot\partial{\bf x}/\partial\varphi$ denotes the single-particle toroidal canonical angular momentum. 

Lastly, we note that Eq.~\eqref{eq:P_phi_dot}  could have been obtained directly from the guiding-center Noether equation \eqref{eq:Noether_gc}, with $P_{\varphi} \equiv \delta{\cal S}/\delta\varphi$ obtained from Eq.~\eqref{eq:deltaS_xt}, and 
\[ \left( \fd{\Phi}{\varphi}, \fd{\bf A}{\varphi} \right) = \left( \frac{1}{c}\,\pd{\psi}{t} - \frac{1}{c}\,\pd{\psi}{t},\; \pd{\bf x}{\varphi}\btimes{\bf B} + \nabla\psi \right) \equiv 0 \]
obtained from Eq.~\eqref{eq:delta_PhiA_xt}, with $\psi \equiv -\,\delta\chi/\delta\varphi$.

\section{\label{sec:EP}Euler-Poincar\'{e} Variational Principle}

In this last Section, we present the Euler-Poincar\'{e} variational principle for the guiding-center Vlasov-Maxwell equations derived from the Lagrange variational principle \eqref{eq:Agc_Lag}.  The Euler-Poincar\'e approach is based on a reduction process \cite{HoMaRa} that exploits the geometric relation between Lagrangian and Eulerian variables to take a Lagrange variational principle into another one (here called ``Euler-Poincar\'e variational principle'') that is entirely expressed in terms of Eulerian variables. This approach was first developed in plasma kinetic theory by Cendra {\it et al.} \cite{CHHM_1998}. Recently, this has been used in Refs.~\cite{Tronci,TrCa} to formulate new quasi-neutral kinetic models. Its adaptation to gyrocenter theories is found in Ref.~\cite{Squire_Qin_Tang_Chandre2013}. 

\subsection{Eulerian and Lagrangian representations}

We now briefly review the connection between the Lagrangian and Eulerian representations of guiding-center phase-space dynamics \cite{HoMaRa}. First, we denote the Lagrangian path ${\bf z}(t;{\bf z}_{0})$ in terms of an orbit generated the Lie-group action of a left translation  ${\sf g}^{t}: {\bf z}(t;{\bf z}_{0}) \equiv {\sf g}^{t}({\bf z}_{0})$ and 
$({\sf g}^{t})^{-1} \equiv {\sf g}^{-t}$ so that the group action is reversible ${\sf g}^{-t}({\bf z})=({\sf g}^{-t}\circ {\sf g}^t)({\bf z}_0)\equiv {\bf z}_{0}$. Here, the Lie group is given by smooth invertible one-parameter transformations from phase-space to itself, while the corresponding   group operation is denoted by  $\circ$, as it coincides with the usual composition of functions. Notice that we are not enforcing these transformations to be canonical: as we shall see, we can recover this property {\it a posteriori}. 

Next, we calculate the Lagrangian velocity $\partial{\bf z}/\partial t = (d{\sf g}^{t}/dt)({\bf z}_{0}) \equiv \dot{\sf g}^{t}({\bf z}_{0})$, which is expressed in terms of the Lagrangian position 
${\bf z}$ as
\begin{equation}
\pd{\bf z}{t} \;=\; \dot{\sf g}^{t}({\bf z}_{0}) \;=\; \dot{\sf g}^{t}\left({\sf g}^{-t}({\bf z})\right) \;\equiv\;  \left(\dot{\sf g}^{t}\circ{\sf g}^{-t}\right)({\bf z}).
\label{eq:z_dot}
\end{equation}
In order to take the Lagrange action \eqref{eq:Agc_Lag} into the  Euler-Poincar\'e form, it is customary to introduce an Eulerian phase-space ``velocity'' field $\Xi^{a}(\vb{\zeta},t;\mu)$ at the Eulerian position $\vb{\zeta}=(\boldsymbol{\sf X},{\sf p}_\parallel)$:
\begin{equation}
\left(\dot{\sf g}^{t}\circ{\sf g}^{-t}\right)(\vb{\zeta}) \;\equiv\; \vb{\Xi}(\vb{\zeta},t) = \left({\bf U}(\vb{\zeta},t),\frac{}{} \phi_{\|}(\vb{\zeta},t)\right),
\label{phasespacevelocity}
\end{equation}
where the $\mu$-dependence has been omitted for convenience. We also introduce the Eulerian guiding-center Vlasov phase-space density $F_{\mu}(\vb{\zeta},t)$ by a push-forward operation ${\sf T}_{t}^{-1}$, which is defined as
\[ f({\bf z}) \equiv {\sf T}_{t}^{-1}f_{0}({\bf z}) \equiv f_{0}\big({\sf g}^{-t}({\bf z})\big) = f_{0}({\bf z}_{0}), \]
so that its evaluation along a Lagrangian path yields:
\begin{equation}
F_{\mu}\,d^{4}\zeta \;=\; {\sf T}_{t}^{-1}\left(F_0\frac{}{}d^{4}z_{0}\right).
\label{particleconservation}
\end{equation} 

With these definitions, the guiding-center Lagrange functional associated to the action ${\cal A}_{\rm gc}^{\rm L} $ in \eqref{eq:Agc_Lag} is defined in the following Euler-Poincar\'e form:
 \begin{eqnarray}
 {\cal A}_{\rm gc}^{\rm EP} & = & \int_{t_{1}}^{t_{2}}\; {\rm L}_{\rm gc}^{\rm EP}(\vb{\Xi}, F_{\mu},\Phi,{\bf A},{\bf E},{\bf B}) \;dt \nonumber \\
 & \equiv &  \int F_{\mu}(\vb{\zeta}) \left( \frac{e}{c}\,{\bf A}^{*}(\vb{\zeta})\bdot\mathbf{U}(\vb{\zeta}) -  H_{\rm gc}(\vb{\zeta}) \right) d^4\zeta\,d\mu\;dt \nonumber    \\
   &  &+\; \int\! \left( |{\bf E}(\mathbf{x})|^{2} \;-\frac{}{} |{\bf B}(\mathbf{x})|^{2} \right)\;\frac{d^3x\,dt}{8\pi},
  \label{eq:Agc_EP}  
  \end{eqnarray}
where $H_{\rm gc}(\vb{\zeta}) ={{\sf p}_{\|}^{2}}/{2m} + e\,\Phi^{*}({\vb{\sf X}},\mu)$ and the explicit dependence on time has been omitted for convenience. 
In the Euler-Poincar\'{e} Lagrangian ${\rm L}_{\rm gc}^{\rm EP}$ defined in Eq.~\eqref{eq:Agc_EP}, the variational fields are the guiding-center Vlasov-Maxwell fields $(F_{\mu},\Phi,{\bf A},{\bf E},{\bf B})$, as well as  the Eulerian velocities $\Xi^{a}$ defined in Eq.~\eqref{phasespacevelocity}.

We now show how the guiding-center Euler-Poincar\'{e} variational principle
\begin{equation}
\delta\int_{t_{1}}^{t_{2}}  {\rm L}_{\rm gc}^{\rm EP}(\vb{\Xi},F_{\mu},\Phi,{\bf A},{\bf E},{\bf B})\,dt \;\equiv\; 0
\label{eq:action_EP}
\end{equation}
can be used to derive the guiding-center Vlasov-Maxwell equations in Eulerian form. 

\subsection{Euler-Poincar\'e variations}

In accordance with the Euler-Poincar\'{e} formalism \cite{CHHM_1998}, the Eulerian variations $(\delta F_\mu, \delta\Xi^a)$ of the guiding-center phase-space fields 
$(F_{\mu}, \Xi^{a})$ are generated by the (Eulerian) virtual displacements in guiding-center phase space. First, we introduce a Lagrangian-path perturbation Lie-group action generated by the left translation $\wt{\sf g}_{\epsilon}$, such that the perturbed Lagrangian path is $\wt{\bf z} \equiv \wt{\sf g}_{\epsilon}({\bf z}_0)$. Notice that $\wt{\sf g}_{\epsilon}$ is obtained by deforming the whole path ${\sf g}^t$ and, therefore, the new translation retains its time dependence (so that  $\wt{\sf g}_{0}({\bf z}_0) ={\sf g}^t({\bf z}_0)= {\bf z}$), although, here, we avoid the notation $\wt{\sf g}_{\epsilon}^t$ for convenience. 

Next, we calculate $\partial\wt{\bf z}/\partial\epsilon =(d\wt{\sf g}_{\epsilon}/d\epsilon)( {\bf z}_0) \equiv \wt{\sf g}_{\epsilon}^{\prime}({\bf z}_0)$, which can be expressed in terms of 
$\wt{\bf z}$ as
\begin{equation} 
\pd{\wt{\bf z}}{\epsilon} \;=\; \wt{\sf g}_{\epsilon}^{\prime}( {\bf z}_0) \;\equiv\; \wt{\sf g}_{\epsilon}^{\prime}\left(\wt{\sf g}_{\epsilon}^{-1}(\wt{\bf z})\right) \;=\; 
\left(\wt{\sf g}_{\epsilon}^{\prime}\circ\wt{\sf g}_{\epsilon}^{-1}\right)(\wt{\bf z}),
\label{eq:z_epsilon}
\end{equation}
according to a definition analogous to Eq.~\eqref{eq:z_dot}. Hence, in order to calculate the Eulerian variation $\delta\vb{\Xi}$ of Eq.~\eqref{phasespacevelocity}, we define the Lagrangian-path variation as $\delta{\bf z} \equiv (\partial\wt{\bf z}/\partial\epsilon)_{\epsilon = 0} = \wt{\sf g}_{\epsilon}^{\prime}({\bf z})\big|_{\epsilon = 0}$,
which can be expressed in Eulerian form as
 \begin{equation}
 \left(\wt{\sf g}_{\epsilon}^{\prime}\circ\wt{\sf g}_{\epsilon}^{-1}\right)(\vb{\zeta})\big|_{\epsilon = 0} \equiv \Delta^{a}(\vb{\zeta},t) = \left(\vb{\xi}(\boldsymbol{\zeta},t),\frac{}{}\kappa_{\|}(\boldsymbol{\zeta},t)\right).
 \label{eq:delta_z_Euler}
 \end{equation}
The Eulerian variation of $F_{\mu}$ is, thus, expressed as
\begin{equation}
\delta F_{\mu}  \;=\; -\;\pd{}{\zeta^{a}}\left(\Delta^{a}\frac{}{} F_{\mu}\right),
\label{eq:deltaF_EP}
\end{equation}
which follows from the fact that the Vlasov phase-space density $F_{\mu}$ must satisfy the particle-number conservation law \eqref{particleconservation}. The Eulerian variation of Eq.~\eqref{phasespacevelocity} is now defined as 
\[
\delta{\boldsymbol\Xi}=\frac{\partial}{\partial \epsilon}\!\left( \dot{\wt{\sf g}}_{\epsilon}\circ \wt{\sf g}_{\epsilon}^{-1}\right)\bigg|_{\epsilon=0}
\]
and is found in Appendix \ref{ref:appendix} to be given by
\begin{equation}
\delta\vb{\Xi} \;-\; \pd{\vb{\Delta}}{t} \;=\; \vb{\Xi}\cdot\pd{\vb{\Delta}}{\vb{\zeta}} \;-\; \vb{\Delta}\cdot\pd{\vb{\Xi}}{\vb{\zeta}} \;\equiv\; \left[\vb{\Xi},\frac{}{} \vb{\Delta}\right],
\label{eq:Lie_bracket}
\end{equation}
where $[\cdot\,,\cdot]$ denotes the Jacobi-Lie bracket of the two vector fields \eqref{phasespacevelocity} and \eqref{eq:delta_z_Euler}. When Eq.~\eqref{eq:Lie_bracket} is written in component form, we find
\begin{equation}
\delta \Xi^{a} \;=\; \pd{\Delta^{a}}{t} \;+\; \Xi^{b}\;\pd{\Delta^{a}}{\zeta^{b}} \;-\; \Delta^{b}\;\pd{\Xi^{a}}{\zeta^{b}}. \label{eq:delta_Xi_EP}
\end{equation}
We note that the variations \eqref{eq:delta_Xi_EP} follow from the relation \eqref{phasespacevelocity}, which coincide with the standard Euler-Poincar\'{e}  form \cite{CHHM_1998} for the variation of vector fields by Lie-transform method:
\begin{eqnarray}
\delta \Xi^{a} & \equiv & \frac{d}{d\epsilon}\left[ {\sf T}_{\epsilon}^{-1}\;\left(\frac{d({\sf T}_{\epsilon}\;\zeta^{a})}{dt}  \right) \right]_{\epsilon = 0} \nonumber \\
 & = & \frac{d}{dt}\left[\frac{d({\sf T}_{\epsilon}\zeta^{a})}{d\epsilon}\right]_{\epsilon = 0} \;+\; \left[ \frac{d({\sf T}_{\epsilon}^{-1}\Xi^{a})}{d\epsilon}\right]_{\epsilon = 0} \nonumber \\
  & = & \frac{d\Delta^{a}}{dt} \;-\; \Delta^{b}\;\pd{\Xi^{a}}{\zeta^{b}},
 \end{eqnarray}
where $d/dt \equiv \partial/\partial t + \Xi^{b}\,\partial/\partial\zeta^{b}$ and $d/d\epsilon \equiv \partial/\partial \epsilon + \Delta^{b}\,\partial/\partial\zeta^{b}$, as well as
${\sf T}_{\epsilon}^{\pm 1} $ denote the push-forward ${\sf T}_{\epsilon}^{-1}$ and pull-back ${\sf T}_{\epsilon}$ operators associated with the guiding-center phase-space displacement \eqref{eq:delta_z_Euler}. 

\subsection{Euler-Poincar\'e equations}

Upon introducing the Eulerian canonical momentum $\boldsymbol{\sf P}=(e/c){\bf A}(\vb{\sf X},t) + {\sf p}_{\|}\bhat(\vb{\sf X},t)$, 
the variation of the guiding-center Euler-Poincar\'{e} Lagrangian  \eqref{eq:Agc_EP}   yields the expression
\begin{eqnarray}
\delta{\rm L}_{\rm gc}^{\rm EP} & = & \int \left[ \delta F_{\mu} \left( \boldsymbol{\sf P}\bdot{\bf U} \;-\frac{}{} H_{\rm gc}\right) \;+\; F_{\mu}\boldsymbol{\sf P}\bdot\delta{\bf U} \right. \nonumber \\
 &  &\left.\hspace*{0.3in}+\; F_{\mu} \left(\delta\boldsymbol{\sf P}\bdot{\bf U} \;-\frac{}{} e\,\delta\Phi \;-\; \mu\,\delta B \right)\right] d^{4}\zeta\,d\mu \nonumber \\
 &  &+\; \int \left( \delta{\bf E}\bdot{\bf E} \;-\frac{}{} \delta{\bf B}\bdot{\bf B} \right)\;\frac{d^{3}x}{4\pi} ,
 \label{eq:deltaL_EP_prime}
\end{eqnarray}
where  $\delta\boldsymbol{\sf P} = (e/c)\;\delta{\bf A} + {\sf p}_{\|}\;\delta{\bf B}\bdot\partial\bhat/\partial{\bf B}$. Upon recalling the definition \eqref{eq:Hgc_field}, the last two lines of Eq.~\eqref{eq:deltaL_EP_prime} can be expressed in terms of the potential variations $(\delta\Phi,\delta{\bf A})$ as
\begin{eqnarray}
\delta{\rm L}_{\rm gc}^{\rm EP} & = & \int \left[ \delta F_{\mu} \left( \boldsymbol{\sf P}\bdot{\bf U} \;-\frac{}{} H_{\rm gc}\right) \;+\; F_{\mu}\boldsymbol{\sf P}\bdot\delta{\bf U} \right] 
d^{4}\zeta\,d\mu \nonumber \\
 & + & \frac{\delta\Phi}{4\pi} \left( \nabla\bdot{\bf E} \;-\frac{}{} 4\pi\int \,e\;F_{\mu}\,d{\sf p}_{\|}\,d\mu\right)  \label{eq:deltaL_EP_2} \\
  & + & \frac{\delta{\bf A}}{4\pi}\bdot\left( \frac{1}{c}\pd{\bf E}{t} + \frac{4\pi}{c}\int  e\,{\bf U}\;F_{\mu}\,d^{4}\zeta\,d\mu - \nabla\btimes{\bf H} \right),
\nonumber 
\end{eqnarray}
so that stationarity with respect to the potential variations $(\delta\Phi,\delta{\bf A})$ yields the guiding-center Maxwell equations \eqref{eq:div_E}-\eqref{eq:curl_B}, respectively.

The remaining terms in the variation of the guiding-center Euler-Poincar\'{e} Lagrangian \eqref{eq:deltaL_EP_2} are
\begin{eqnarray}
\delta{\rm L}_{\rm gc}^{\rm EP} & = & \int \left[ \delta F_{\mu} \left( \boldsymbol{\sf P}\bdot{\bf U} \;-\frac{}{} H_{\rm gc}\right) \;+\; F_{\mu}\boldsymbol{\sf P}\bdot\delta{\bf U}\right] d^{4}\zeta
\,d\mu \nonumber \\  
 & = & \int  F_{\mu} \left[\vb{\xi}\bdot\left( e\,{\bf E}^{*} \;+\; \frac{e}{c}\,{\bf U}\btimes{\bf B}^{*} \;-\; \phi_{\|}\,\bhat \right) \right. \nonumber \\
  &  &\left.\hspace*{0.5in}+\; \kappa_{\|} \left( \bhat\bdot{\bf U} - \frac{{\sf p}_{\|}}{m} \right) \right] d^{4}\zeta\,d\mu,
\label{eq:deltaL_EP}
\end{eqnarray}
where $\delta F_{\mu}$ and $\delta{\bf U}$ are given by Eqs.~\eqref{eq:deltaF_EP}-\eqref{eq:delta_Xi_EP}. Hence, the stationarity of the guiding-center 
Euler-Poincar\'{e} action \eqref{eq:action_EP} with respect to arbitrary variations $(\vb{\xi},\kappa_{\|})$ yields (after some rearrangements)
\[
{\mathbf{U}}
=
\frac1{B^*_\parallel}\left(\frac{{\sf p}_\parallel}m\,\mathbf{B}^*+c\mathbf{E}^*\times\bhat\right)
,\qquad\,
\phi_\parallel
=
\frac{e}{B^*_\parallel}\mathbf{E}^*\cdot\mathbf{B}^*,
\]
which are the Eulerian analogs of the guiding-center Euler-Lagrange equations \eqref{eq:EL_X}-\eqref{eq:EL_p} in Lagrangian coordinates. Hence, the Vlasov equation for $F_{\mu}$ is  obtained by replacing the expression above of the phase-space vector  $\Xi^a$ into the transport equation 
$\partial_t F_{\mu} + \partial(\Xi^{a} F_{\mu})/\partial\zeta^{a}=0$, which is obtained by taking the time derivative of the particle conservation relation \eqref{particleconservation}.

\subsection{Noether theorem and the symplectic form}

The Euler-Poincar\'e approach for collisionless kinetic equations described in this Section is similar in construction to the Euler-Poincar\'e reduction for fluid flows \cite{HoMaRa} (recall that the Lagrangian phase-space coordinates carry the parametric dependance on the magnetic moment). In this context, the method exploits the relabeling symmetry to obtain (by Noether's theorem) the conservation of the Kelvin circulation of the fluid velocity. Although, the Lagrangian  in Eq.~\eqref{eq:Agc_Lag} is not generally invariant under relabeling (which is why the Eulerian density $F_{\mu}$ appears as a new dynamical variable), a symmetry is still present and it is associated to relabeling (volume-preserving) transformations of phase-space that preserve the reference density $F_0$. 

Indeed, if $\bar{\sf g}^{\tau}$ is any one-parameter volume-preserving transformation, i.e., the volume form $F_0\,d^{4}z_{0}$ is preserved so that Eq.~\eqref{particleconservation} reads $\ov{\sf T}_{\tau}^{-1}(F_0\,d^{4}z_{0}) = F_0\,d^{4}z_{0}$, then the Lagrangian in Eq.~\eqref{eq:Agc_Lag} is left fully invariant under the relabeling action of $\bar{\sf g}^{\tau}$, that is ${L}_{\rm gc}({\sf g}^t,\dot{{\sf g}}^t)={L}_{\rm gc}({\sf g}^t\circ\bar{\sf g}^{\tau},\dot{{\sf g}}^t\circ\bar{\sf g}^{\tau})$, where  $ {L}_{\rm gc}({\sf g}^t,\dot{{\sf g}}^t)$ is the Lagrangian obtained by replacing ${\bf z}={\sf g}^t({\bf z}_0)$ in \eqref{eq:Lgc_Lag} (here, the dependence on the field variables has been omitted for convenience). At this point, one can look for a Noether conserved quantity associated to this symmetry \cite{HoMaRa,CHHM_1998}. The standard Noether method was recently applied to the Euler-Poincar\'e construction in Ref.~\cite{Holm} (see Corollary 9.1.2 therein). As we shall see, the Noether conserved quantity naturally unfolds the symplectic features of the phase-space flow in both Lagrangian and Eulerian coordinates. Indeed, in the Euler-Poincar\'e approach on phase-space, no symplectic Hamiltonian properties are assumed \emph{a priori} in the reduction from Lagrangian to Eulerian coordinates (for example, the Lagrangian flow ${{\sf g}}^t$ was never assumed to be canonical). Rather, these properties emerge as a consequence of the specific symmetry possessed by the Lagrangian \eqref{eq:Lgc_Lag}, whose Noether conserved quantity returns precisely preservation of the guiding-center symplectic form.

In the present case, the Noether conserved quantity can be found by noting that, in order to obtain Eq.~\eqref{eq:deltaL_EP}, the following boundary term has been set to zero:
\begin{multline}
 \int \!\bigg(\frac{\delta{\rm L}_{\rm gc}^{\rm EP}}{\delta U^{i}}\,\xi^{i}+\frac{\delta{\rm L}_{\rm gc}^{\rm EP}}{\delta \phi_\parallel}\,\kappa_\parallel\bigg)\,d{\sf p}_{\|}d\mu 
 \,\bigg|_{t_1}^{t_2}  \\
 =\; \int \!F_{\mu}\;\boldsymbol{\sf P}\cdot\vb{\xi}\,d{\sf p}_{\|}d\mu \,\bigg|_{t_1}^{t_2} =\, 0,
\end{multline}
because Eq.~\eqref{eq:delta_z_Euler} vanish at the endpoints. This calculation shows how the Eulerian one-form 
${\sf P}_{i\,}\exd{\sf X}^i$ emerges in the tensor contraction $\boldsymbol{\sf P}\cdot\vb{\xi}$ against the vector field ${\xi^i}\partial/\partial{\sf X}^i$.

\rem{ 
Now, recall \eqref{lagrangianvariables}: since $\delta \mathbf{X} \equiv \vb{\xi}(\boldsymbol{\zeta},t;\mu)$, we have 
\[
\vb{\xi}(\vb{\zeta},t;\mu)=(\mathsf{T}^{-1}_{{\bf z}}\delta\mathbf{X})(\vb{\zeta},t;\mu)
=(\delta\mathbf{X}\circ{\bf z}^{-1})(\vb{\zeta},t;\mu)
\]
and thus
\begin{align*}
\int\!d^{4}\zeta\,d\mu\, F_{\mu}\;\boldsymbol{\sf P}\cdot\mathsf{T}^{-1}_{{\bf z}}\delta\mathbf{X}=&
\int \!d^{4}z_0\,d\mu\,\mathsf{T}_{{\bf z}}(F\boldsymbol{\sf P})\cdot\delta\mathbf{X}
\\
=&\int \!d^{4}z_0\,d\mu\,F_0\mathsf{T}_{{\bf z}}\boldsymbol{\sf P}\cdot\delta\mathbf{X}
\end{align*}
where the last equality follows by the distributive property of the pullback action on tensor fields and the definition \eqref{particleconservation}. Here, $\mathsf{T}_{{\bf z}}\boldsymbol{\sf P}\cdot\delta\mathbf{X}$ stands for the contraction between the tangent vector $\delta\mathbf{X}$ and the  pull-back by ${\bf z}$ of the Eulerian one-form $\boldsymbol{\sf P}\cdot{\sf d}\vb{\sf X}$
}   

At this point, standard application of the Noether method would require writing the Eulerian displacement $\vb{\xi}=\delta\mathbf{X}\circ{\sf g}^{-t}$, as it is expressed in terms of Lagrangian variables \eqref{eq:Lag_z}, as an infinitesimal generator of volume-preserving transformations, thereby leading to the conservation law 
\begin{equation}
{\sf d}\big[{\sf T}_{t}{(\boldsymbol{\sf P}\cdot{\sf d}\vb{\sf X})}\big]={\sf d}(\boldsymbol{\sf P}\cdot{\sf d}\vb{\sf X}).
\label{KNcons}
\end{equation}
For the sake of simplicity, here we shall simply verify this relation by using the Euler-Lagrange Eqs.~\eqref{eq:EL_X}-\eqref{eq:EL_p}. Indeed, after recalling Eq.~\eqref{eq:Lag_z} and computing the pull-back operation ${\sf T}_{t}(\boldsymbol{\sf P}\cdot{\sf d}\vb{\sf X})=\mathbf{P}\cdot{\sf d}\mathbf{X}$, Eqs.~\eqref{eq:EL_X}-\eqref{eq:EL_p} yield the Lagrangian relation
\begin{eqnarray}
\exd\Lambda_{\rm gc} & = & \nabla\Lambda_{\rm gc}\bdot\exd{\bf X} \;+\; \pd{\Lambda_{\rm gc}}{{\sf p}_{\|}}\;\exd {\sf p}_{\|} \nonumber \\
 & = & \frac{d}{dt}\left(\pd{\Lambda_{\rm gc}}{\dot{\bf X}}\right)\bdot\exd{\bf X} \;\equiv\; \frac{d}{dt}\left({\bf P}\bdot\exd{\bf X}\right),
 \end{eqnarray}
 so that Eq.~\eqref{KNcons} is recovered by taking the differential on both sides.
 
If we now use Stokes Theorem ($\int_{\cal S}\exd\alpha = \oint_{\partial{\cal S}}\alpha$ for any differential $k$-form $\alpha$ and any open surface 
${\cal S}$ with closed boundary $\partial{\cal S}$), we find
\begin{equation}
0 \,\equiv\, \oint_{\gamma_0\!}\exd\Lambda_{\rm gc} \,=\, \frac{d}{dt}\oint_{\gamma_0\!}{\bf P}\bdot\exd{\bf X}
\,=\, \frac{d}{dt}\oint_{\gamma(t;\mu)\!}\boldsymbol{\sf P}\bdot\exd\vb{\sf X},
\label{eq:EP_inv}
\end{equation}
where $\gamma_0$ is an arbitrary fixed loop while the closed contour $\gamma(t;\mu) \equiv {\sf g}^{t}(\gamma_{0})$ moves with the Lagrangian flow. Here,  we used 
$\exd^{2}\Lambda_{\rm gc} \equiv 0$ and the last equality follows from relabeling inside the circulation integral, that is $\oint_{\gamma_0}\!{\sf T}_{t}(\boldsymbol{\sf P}\bdot\exd
\vb{\sf X})=\oint_{\,{{\sf g}}^t(\gamma_0)}\boldsymbol{\sf P}\bdot\exd\vb{\sf X}$.

The circulation $\oint_{\gamma(t;\mu)}\boldsymbol{\sf P}\bdot\exd\vb{\sf X}$ in Eq.~\eqref{eq:EP_inv} is the Poincar\'{e} relative integral invariant for guiding-center motion. Application of Stokes Theorem yields $\iint_{\sigma(t;\mu)}\exd {\sf P}_{i}\wedge \exd {\sf X}^{i}=const.$, where $\sigma(t)$ is the surface surrounded by the arbitrary loop $\gamma(t;\mu)$. Since this surface is also arbitrary, the relation
\[
\frac{d}{d t}\iint_{\sigma(t;\mu)}\!\exd\vb{\sf X}\wedge\exd\boldsymbol{\sf P}=0
\]
naturally recovers conservation of the guiding-center symplectic form $\Omega_{\rm gc}={\exd\vb{\sf X}\wedge \exd\boldsymbol{\sf P}}$ [see Eq.~(3.19) in Ref.~\cite{Cary_Brizard_2009}  and the first two terms of Eq.~\eqref{eq:omega_gc} in this paper]. Indeed, Eq.~\eqref{KNcons} yields ${\sf T}_{t}\Omega_{\rm gc}=\Omega_{\rm gc}$, as it arises from commutation of pullback with the differential. Then, the conservation of the Liouville density $mB^*_\parallel (\vb{\sf X})d^3{\sf X}d {\sf p}_\parallel$ [see Eq.~\eqref{eq:gcLiouville}] is a natural consequence of the preservation of $\Omega_{\rm gc}$, as explained in detail in Ref.~\cite{Cary_Brizard_2009}.


Lastly, we note that, while the guiding-center Vlasov-Maxwell equations have been rederived by the guiding-center Euler-Poincar\'{e} variational principle
\eqref{eq:action_EP}, it can also be used to derive the Hamiltonian functional bracket underlying the Hamiltonian structure of the guiding-center Vlasov-Maxwell equations. The guiding-center Hamiltonian functional $\mathbb{H}_{\rm gc}$ is then constructed by Legendre transformation from the
Euler-Poincar\'{e} Lagrangian density \eqref{eq:Agc_EP} as
\begin{widetext}
\begin{eqnarray}
\mathbb{H}_{\rm gc} & = & \int  \left[ \pd{\bf A}{t}\bdot\pd{{\cal L}_{\rm gc}^{\rm EP}}{(\partial_{t}{\bf A})} + \left(\int {\bf U}\bdot
\pd{{\cal L}_{\rm gc}^{\rm EP}}{\bf U} d{\sf p}_{\|}\,d\mu\right) - {\cal L}_{\rm gc}^{\rm EP} \right] d^{3}x \nonumber \\
 & = & \int  F_{\mu}K_{\rm gc}\,d^{4}\zeta\,d\mu+ \int  d^{3}x\left[ \frac{1}{8\pi} \left(|{\bf E}|^{2} \;+\frac{}{} |{\bf B}|^{2} \right) \;+\;
 \nabla\bdot\left(\frac{\bf E}{4\pi}\;\Phi\right) \right] \;\equiv\; \int  {\cal E}_{\rm gc}\;d^{3}x,
\end{eqnarray}
\end{widetext}
where the guiding-center energy density ${\cal E}_{\rm gc}$ was derived previously [see Eq.~\eqref{eq:E_gc}] by Noether method and the divergence term in the Maxwell term integrates away. This approach was followed for the Maxwell-Vlasov system in Ref.~\cite{CHHM_1998} [see Eqs.~(7.2)-(7.3) therein]. 

Next, we need to construct a functional bracket $[\;,\;]_{\rm gc}$ in terms of which the guiding-center Vlasov-Maxwell equations \eqref{eq:gcV_div}, \eqref{eq:curl_E}, and \eqref{eq:curl_B} are expressed in Hamiltonian functional form as
\begin{eqnarray}
\dot{\mathbb{F}} \equiv \left[ \mathbb{F},\frac{}{} \mathbb{H}_{\rm gc} \right]_{\rm gc} & = & \int  \pd{F_{\mu}}{t}\;\fd{\mathbb{F}}{F_{\mu}}\,d^{4}\zeta\,d\mu 
\label{eq:PBgc_func} \\
 &  &+\; \int  \left(\pd{\bf E}{t}\bdot\fd{\mathbb{F}}{\bf E} + \pd{\bf B}{t}\bdot\fd{\mathbb{F}}{\bf B}\right) d^{3}x,
 \nonumber
\end{eqnarray}
where $\mathbb{F}[F_{\mu},{\bf E},{\bf B}]$ is an arbitrary functional of the guiding-center fields $(F_{\mu},{\bf E},{\bf B})$. While this last step is complicated by the degeneracy of the Euler-Poincar\'e Lagrangian ${\rm L}_{\rm gc}^{\rm EP}$ (as already noticed in Ref.~\cite{CHHM_1998} for the Maxwell-Vlasov case), these complications can be overcome by the use of Dirac constraints, as presented in Ref.~\cite{Squire_Qin_Tang_Chandre2013}. The remaining Maxwell equations \eqref{eq:div_B} and \eqref{eq:div_E} are treated as initial conditions. The construction of the functional bracket $[\;,\;]_{\rm gc}$ is, however, outside the scope of the present paper and will be investigated in a future paper.

\section{Summary}

In this paper, we have presented three different variational formulations for the guiding-center Vlasov-Maxwell equations \eqref{eq:gcV} and \eqref{eq:div_E}-\eqref{eq:curl_B}: the guiding-center Lagrange action functional \eqref{eq:Agc_Lag}, the guiding-center Euler action functional \eqref{eq:Agc_Eul}, and the guiding-center Euler-Poincar\'{e} action functional \eqref{eq:Agc_EP}. Each guiding-center variational principle $\delta{\cal A}_{\rm gc} = 0$, which yields its own version of the guiding-center Vlasov-Maxwell equations, included a self-consistent derivation of the guiding-center magnetization \eqref{eq:Mgc}, expressed in terms of its intrinsic magnetic-dipole and moving electric-dipole contributions 
\eqref{eq:mu_gc}-\eqref{eq:pi_gc}, respectively, as shown in Eqs.~\eqref{eq:Lgc_B},  \eqref{eq:PBgc_B}, or \eqref{eq:deltaL_EP_prime}. Here, the moving electric-dipole contribution \eqref{eq:pi_gc} is expressed in terms of the guiding-center polarization \cite{Brizard_2013,Tronko_Brizard_2015}, which involves the perpendicular magnetic-drift velocity.

One of the great advantages of a variational formulation of the guiding-center Vlasov-Maxwell equations is that their exact conservation laws can be derived directly by Noether method. In Sec.~\ref{sec:Noether}, we used the guiding-center Noether equation \eqref{eq:Noether_gc} to derive the guiding-center energy-momentum conservation laws as well as the conservation law of guiding-center toroidal canonical angular momentum derived for axisymmetric tokamak geometry. In each guiding-center conservation law, we showed the crucial played by the complete form of guiding-center magnetization \eqref{eq:Mgc}. We also showed that, thanks to the contribution of the guiding-center polarization
\eqref{eq:pi_gc}, the guiding-center stress tensor \eqref{eq:Tgc_def} was explicitly shown to be symmetric, as required by the conservation of guiding-center canonical angular momentum. In contrast, previous works \cite{Similon_1985} have only implicitly assumed the symmetry of the guiding-center stress tensor, based on the requirements of the conservation of guiding-center canonical angular momentum.

Lastly, future work will involve the construction of a guiding-center Hamiltonian field theory based on the explicit construction of the guiding-center functional bracket $[\;,\;]_{\rm gc}$ used in the guiding-center Hamiltonian functional equation \eqref{eq:PBgc_func}.

Work by AJB was supported by a U.~S.~Dept.~of Energy grant under contract No.~DE-SC0006721. CT acknowledges financial support by the Leverhulme Trust Research Project Grant 2014-112, and by the London Mathematical Society Grant No. 31320 (Applied Geometric Mechanics Network).
    
\appendix

\section{Variational relations\label{ref:appendix}}

This Appendix proves the variational formula \eqref{eq:Lie_bracket} as follows. Starting from the definition $\boldsymbol\Xi= \dot{\sf g}^{t}\circ {\sf g}^{-t}$, we apply the chain rule repeatedly to write 
\begin{eqnarray*}
\delta\vb{\Xi} &  \equiv & \delta(\dot{\sf g}\circ {\sf g}^{-1}) \;=\; \delta\dot{\sf g}\circ {\sf g}^{-1} \;+\; \nabla\dot{\sf g}\cdot \delta{\sf g}^{-1} \\
 & = & \frac{\partial}{\partial t}\big(\delta{\sf g}\circ {\sf g}^{-1}\big) \;-\; \nabla\delta{\sf g}\cdot\frac{\partial}{\partial t}\!\left({\sf g}^{-1}\right) \;+\; \nabla\dot{\sf g}\cdot \delta{\sf g}^{-1} \\
  & = & \frac{\partial\boldsymbol\Delta}{\partial t} + \nabla\delta{\sf g}\cdot\left[(\nabla{\sf g})^{-1}\cdot\boldsymbol\Xi\right] - \nabla\dot{\sf g}\cdot \left[(\nabla{\sf g})^{-1}\cdot\boldsymbol\Delta\right]
\end{eqnarray*}
where we have dropped the superscript $t$ for convenience and we have used the notation $\nabla=\nabla_{\!{\sf g}^{-1}}$, as well as the identities
\begin{align*}
\delta{\sf g}^{-1}=&\ -(\nabla{\sf g})^{-1}\cdot(\delta{\sf g}\circ{\sf g}^{-1}) \equiv -(\nabla{\sf g})^{-1}\cdot\boldsymbol\Delta
\\
{\partial_t}({\sf g}^{-1})=&\ -(\nabla{\sf g})^{-1}\cdot(\dot{\sf g}\circ{\sf g}^{-1}) \equiv -(\nabla{\sf g})^{-1}\cdot\boldsymbol\Xi
\,,
\end{align*}
as they emerge from $\delta({\sf g}\circ{\sf g}^{-1})={\partial_t}({\sf g}\circ{\sf g}^{-1})=0$. 

Now, as a general property of Jacobian matrices, we have $(\nabla_{\!{\sf g}^{-1\!}({\bf z})}{\sf g})^{-1}=\nabla_{\!\bf z\,}{\sf g}^{-1}$, so that 
\begin{align*}
(\nabla\delta{\sf g})\cdot (\nabla{\sf g})^{-1}=&\ 
\nabla(\delta{\sf g}\circ{\sf g}^{-1})=\nabla\boldsymbol\Delta
\\
(\nabla\dot{\sf g})\cdot (\nabla{\sf g})^{-1}=&\ 
\nabla(\dot{\sf g}\circ{\sf g}^{-1})=\nabla\boldsymbol\Xi
\,,
\end{align*}
which then proves Eq.~\eqref{eq:Lie_bracket}.


\begin{thebibliography}{99}
\bibitem{footnote2} Since the gyroangle $\theta$ is an ignorable coordinate in guiding-center Vlasov-Maxwell theory, the integration over the gyroangle $\theta$ may be replaced with $2\pi$, which is added to the guiding-center Jacobian $m\,B_{\|}^{*}$.
\bibitem{footnote3} The gyromotion term $(mc/e)\,\mu\;\exd\theta$ is not needed in Eq.~\eqref{eq:Gamma_gc}.
\end{thebibliography}
\end{document}